\documentclass[fleqn,usenatbib]{mnras}

\usepackage{newtxtext,newtxmath}

\usepackage[T1]{fontenc}

\DeclareRobustCommand{\VAN}[3]{#2}
\let\VANthebibliography\thebibliography
\def\thebibliography{\DeclareRobustCommand{\VAN}[3]{##3}\VANthebibliography}

\usepackage{graphicx}	
\usepackage{amsmath}	
\usepackage{booktabs}
\usepackage{multirow}
\usepackage[flushleft]{threeparttable}
\usepackage{orcidlink}

\title[TPCF of Star Clusters]{The Dependence of the Hierarchical Distribution of Star Clusters on Galactic Environment}

\author[S.~H.~Menon et al.]{
Shyam~H.~Menon\orcidlink{0000-0001-5944-291X},$^{1}$\thanks{E-mail: shyam.menon@anu.edu.au}
Kathryn~Grasha\orcidlink{0000-0002-3247-5321},$^{1,2}$
Bruce~G.~Elmegreen\orcidlink{0000-0002-1723-6330},$^{3}$
Christoph~Federrath\orcidlink{0000-0002-0706-2306},$^{1,2}$
\newauthor
Mark~R.~Krumholz\orcidlink{0000-0003-3893-854X},$^{1,2}$
Daniela~Calzetti\orcidlink{0000-0002-5189-8004},$^{4}$
N\'estor~S\'anchez\orcidlink{0000-0002-0042-3180},$^{5}$
Sean~T.~Linden\orcidlink{0000-0002-1000-6081},$^{4}$
Angela~Adamo\orcidlink{0000-0002-8192-8091},$^{6}$
\newauthor
Matteo~Messa\orcidlink{0000-0003-1427-2456},$^{4,7}$
David~O.~Cook\orcidlink{0000-0002-6877-7655},$^{8,9}$
Daniel~A.~Dale\orcidlink{0000-0002-5782-9093},$^{10}$
Eva~K.~Grebel\orcidlink{0000-0002-1891-3794},$^{11}$
Michele~Fumagalli\orcidlink{0000-0001-6676-3842},$^{12,13}$
\newauthor
Elena~Sabbi\orcidlink{0000-0003-2954-7643},$^{14}$
Kelsey~E.~Johnson\orcidlink{0000-0001-8348-2671},$^{15}$
Linda~J.~Smith\orcidlink{0000-0002-0806-168X},$^{14}$
Robert~C.~Kennicutt\orcidlink{0000-0001-5448-1821},$^{16,17}$
\\
$^{1}$Research School of Astronomy and Astrophysics, Australian National University, Canberra, ACT~2611, Australia\\
$^{2}$ARC Centre of Excellence for All Sky Astrophysics in 3 Dimensions (ASTRO 3D), Australia\\
$^{3}$IBM Research Division, T.J Watson Research Center, Yorktown Hts, NY~10598, USA\\
$^{4}$Department of Astronomy, University of Massachusetts, Amherst, MA~01003, USA\\
$^{5}$Universidad Internacional de Valencia (VIU), C/Pintor Sorolla~21, E-46002 Valencia, Spain\\
$^{6}$The Oskar Klein Centre, Department of Astronomy, Stockholm University, AlbaNova, SE-10691 Stockholm, Sweden\\
$^{7}$Observatoire de Gen\`{e}ve, Universit\'{e} de Gen\`{e}ve, Chemin Pegasi, 51 1290 Versoix, Switzerland\\
$^{8}$Department of Physics \& Astronomy, California Institute of Technology, Pasadena, CA~91101, USA\\
$^{9}$IPAC/Caltech, Pasadena, CA~91101, USA\\
$^{10}$Department of Physics \& Astronomy, University of Wyoming, Laramie, WY~82071, USA\\
$^{11}$Astronomisches Rechen-Institut, Zentrum f\"ur Astronomie 
der Universit\"at Heidelberg, M\"onchhofstr.\ 12--14, 69120 Heidelberg, 
Germany\\
$^{12}$Dipartimento di Fisica G. Occhialini, Universit\`{a} degli Studi di Milano Bicocca, Piazza della Scienza 3, 20126~Milano, Italy\\
$^{13}$INAF Osservatorio Astronomico di Trieste, via G. Tiepolo~11, Trieste, Italy\\
$^{14}$Space Telescope Science Institute, 3700 San Martin Drive, Baltimore, MD~21218, USA\\
$^{15}$Department of Astronomy, University of Virginia, Charlottesville, VA~22904, USA\\
$^{16}$Steward Observatory, University of Arizona, 933 North Cherry Avenue, Tucson, AZ~85721, USA\\
$^{17}$George~P. and Cynthia~W. Mitchell Institute for Fundamental Physics \& Astronomy, Texas A\&M University, College Station, TX~77843, USA\\
}

\date{Accepted XXX. Received YYY; in original form ZZZ}

\pubyear{2021}

\begin{document}
\label{firstpage}
\pagerange{\pageref{firstpage}--\pageref{lastpage}}
\maketitle

\begin{abstract}
We use the angular Two Point Correlation Function (TPCF) to investigate the hierarchical distribution of young star clusters in 12 local (3--18 Mpc) star-forming galaxies using star cluster catalogues obtained with the \textit{Hubble Space Telescope} (\textit{HST}) as part of the Treasury Program LEGUS (Legacy ExtraGalactic UV Survey). The sample spans a range of different morphological types, allowing us to infer how the physical properties of the galaxy affect the spatial distribution of the clusters. We also prepare a range of physically motivated toy models to compare with and interpret the observed features in the TPCFs. We find that, conforming to earlier studies, young clusters ($T \la 10\, \mathrm{Myr}$) have power-law TPCFs that are characteristic of fractal distributions with a fractal dimension $D_2$, and this scale-free nature extends out to a maximum scale $l_{\mathrm{corr}}$ beyond which the distribution becomes Poissonian. However, $l_{\mathrm{corr}}$, and $D_2$ vary significantly across the sample, and are correlated with a number of host galaxy physical properties, suggesting that there are physical differences in the underlying star cluster distributions. We also find that hierarchical structuring weakens with age, evidenced by flatter TPCFs for older clusters ($T \ga 10\, \mathrm{Myr}$), that eventually converges to the residual correlation expected from a completely random large-scale radial distribution of clusters in the galaxy in $\sim 100 \, \mathrm{Myr}$.  Our study demonstrates that the hierarchical distribution of star clusters evolves with age, and is strongly dependent on the properties of the host galaxy environment.
\end{abstract}




\section{Introduction}
Stars typically do not form in isolation but rather concentrated in clusters \citep{Lada_2003} that carry the imprint of the gas from which stars form \citep{Krumholz_Clusters}. These regions are found in dense, hierarchically structured molecular clouds that accrete gas from their surroundings and undergo gravitational collapse. Young stars and star clusters inherit the spatial properties of the natal gas from which they form, and can therefore be used as tracers to understand the physical mechanisms at play in the star formation cycle. Unlike young stars though, young stellar clusters can be observed to greater distances, and hence provide an excellent source of information to investigate the complex mechanisms of star formation in diverse environments. 

Star formation in galaxies is spatially structured in a hierarchical, scale-free pattern such that, smaller and denser associations that extend all the way down to substellar scales, are surrounded by larger, less dense ones, that go out to kiloparsec scales \citep[see,][for a review]{Elmegreen_2009}. These scale-free structures are analogous to geometric fractals \citep{Mandelbrot_1982}, and have been shown to be present in the distribution of unbound stars \citep[see,][and references therein]{Gouliermis_2018}, embedded stars in clusters and star-forming regions \citep{Sanchez_2007,Fernandes_2012,Gregorio_2015,Sun_2017}, $\ion{H}{ii}$ regions \citep{Feitzinger_1987,Sanchez_2008}, OB associations \citep{Bresolin_1998,Pietrzynski_2001,Kumar_2004,Gutermuth_2008}, star-forming regions \citep{Elmegreen_2001_Fractal,Elmegreen_2006,Bastian_2007,Elmegreen_2014,Rodriguez_2020,Mondal_2021}, and young star clusters \citep{Zhang_2001,Bastian_2005,Scheepmaker_2009,Grasha_2015,Grasha_2017_Spatial,Grasha_2017_Ages}, and are expected to originate from the inherently hierarchically structured interstellar gas distribution \citep{Elmegreen_1999,Elmegreen_2003,Elmegreen_2004,Elmegreen_2007,Bergin_2007,Dutta_2009,Federrath_Fractals,Beattie_2019MachNo}. This hierarchical nature of gas is consistent with that set by the scale-free physical mechanisms that act on it, i.e., gravity and interstellar turbulence, and is central to the so-called gravoturbulent fragmentation \citep{Elmegreen_1993,Klessen_2000,Mac_2004,Padoan_2014,Federrath_2018} and global hierarchical collapse (GHC) \citep{Semadeni_2009,Semadeni_2017} scenarios, both leading theories describing the multi-scale star and cluster formation process in the ISM \citep{Mckee_2007,Krause_2020}. Individual stars form at the smallest scales of this hierarchy and group together to form star clusters, which themselves are spatially correlated with other star clusters in kpc-scale star complexes and flocculent spiral arms \citep{Efremov_1995,Elmegreen_1996,Gusev_2002,Bastian_2005,Ivanov_2005}. The spatial correlations of star clusters would thus trace the largest scales of this hierarchy, that has properties which would presumably be set by the physical mechanisms governing star formation at galactic scales. Studying the spatial structure of star clusters is thus an effective way of obtaining insights into the physical mechanisms at play. 

The two-point correlation function (TPCF) is a robust tool to probe the scale dependence of the clustering properties of a distribution, as it quantifies how much excess correlation a distribution of points has at a given separation (angular or linear) compared to a completely random distribution \citep{Peebles_1980}. For a hierarchical or scale-free distribution, a general trend of the TPCF decreasing with separation is expected \citep{Gomez_1993,Larson_1995,Bate_1998}. Such a trend has been seen in earlier studies of star clusters in galaxies such as the Antennae \citep{Zhang_2001}, M51 \citep{Bastian_2005,Scheepmaker_2009} and NGC~0628 \citep{Grasha_2015}. Apart from providing an estimate for the strength of the correlation at a given scale, the TPCF can also quantify the spatial heterogeneity of the distribution through an effective fractal dimension, that can be obtained from the slope of the TPCF \citep[see, e.g.,][]{Calzetti_1989,Falgarone_1991}. The fractal dimension quantifies how space-filling or clumpy a distribution is, and is expected to be set by turbulence in the ISM \citep{Stutzki_1998,Elmegreen_2004,Sanchez_2005,Federrath_Fractals}. 

It has often been argued that the fractal dimension observed in the ISM has a nearly universal value of around $\sim 2.3$ \citep{Elmegreen_1996_Mass}, suggesting a universal nature of the self-similar hierarchy. However, more recent work has questioned this universality, especially at galactic spatial scales, finding variations in the inferred fractal dimensions and the scales to which the hierarchy extends. Such differences could arise due to the different sources of turbulence that might dominate at various scales and environments, and set different density structures \citep[see;][]{Federrath_Fractals}, and/or from the modification of the scale-free behaviour due to galactic scale dynamical processes such as rotation, shear, or feedback \citep{Padoan_2001,Odekon_2008,Sanchez_2010,Dib_2020}. For instance, \citet{Sanchez_2008} investigated the fractal nature of $\ion{H}{ii}$ regions in 93 nearby galaxies, and found statistically significant variations among the galaxies of the sample, with signs of higher fractal dimensions for brighter, more massive galaxies. Similarly, \citet{Grasha_2017_Spatial}, through the use of the angular TPCF of star clusters in 6 galaxies observed as part of LEGUS (Legacy ExtraGalactic UV Survey), found a large range of fractal dimensions and correlation lengths, with hints at systematic variations with the galaxy stellar mass and star formation rate. If these results are confirmed, it would suggest that the environment of the host galaxy is important in setting the hierarchical structure of star formation at galactic scales, which would qualitatively be consistent with recent evidence for the same in observations \citep[see recent reviews by,][]{Angela_2017,Chevance_2020} and numerical simulations \citep{Kruijssen_2011,Renaud_2018,Pfeffer_2019}. 

The spatial distribution of star clusters is also expected to evolve with age. For instance, there is strong evidence that hierarchical clustering dissipates with age, which manifests as a reduction in the TPCF for older populations of stars and stellar clusters \citep{Bastian_2005,Gieles_2008,Scheepmaker_2009,Sanchez_Alfaro_2009,Sanchez_Alfaro_2010,Gouliermis_2014,Gouliermis_2015,Grasha_2015,Grasha_2017_Spatial}. This decrease in spatial correlation implies that older stars/clusters are more randomly positioned than younger ones. In addition, clusters closer to each other tend to have about the same age, regardless of that age, and this leads to an age-difference versus separation relation in the distribution \citep[see, e.g., ][]{Elmegreen_1996,Efremov_1998,Marcos_2009,Grasha_2017_Ages}. The randomisation of the cluster distribution could be the result of ballistic motion of mutually-unbound clusters away from their birth sites, a product of larger-scale effects such as shear or tidal interactions, or due to the superposition of successive generations of star formation \citep[see, for instance,][for a discussion]{Elmegreen_2018}. Regardless of the mechanism, the observed timescale over which the distribution randomises is \mbox{$\sim 40$--$100 \, \mathrm{Myr}$} \citep{Grasha_2017_Spatial}. However, this timescale also varies from galaxy to galaxy \citep{Grasha_2018,Grasha_2019} and the environment within a galaxy \citep{Silva-Villa_2014,Gouliermis_2015_Review}, and is at least qualitatively consistent with theoretical and numerical work \citep{Elmegreen_2010,Kruijssen_2011,Reina-Campos_2017}. 

In this study, we use the angular TPCF to investigate the environmental dependencies and evolutionary effects of the hierarchical distribution of star clusters in 12 nearby galaxies as part of the Legacy ExtraGalactic UV Survey \citep[LEGUS;][]{Calzetti_2015}. LEGUS is a Cycle~21 \textit{Hubble Space Telescope} (\textit{HST}) Treasury program that imaged 50 nearby (\mbox{$\sim 3$--$18 \, \mathrm{Mpc}$}) galaxies in UV and optical bands, and used the data to identify and prepare catalogues of individual star clusters in the galaxies. We compute and compare the TPCF of the catalogued star clusters among 12~galaxies drawn from the LEGUS sample and search for correlations between the TPCF and the physical conditions of the host galaxy. In addition, we use the estimated ages of the star clusters to probe the time evolution of hierarchical structuring. This study extends the work by \citet{Grasha_2017_Spatial} to 6 more galaxies (12 in total), yielding a larger sample size with more statistical power to constrain any potential dependencies on galaxy properties. In addition, we develop a set of physically motivated toy models to interpret the various features we find in our TPCFs, similar to the approach in \citet{Gouliermis_2014}. 

The paper is outlined as follows: In Section~\ref{sec:Data} we describe the galaxies in our study, and, briefly, the procedure adopted by LEGUS to prepare the star cluster catalogues for them. The methodology we adopt to compute the TPCF is provided in Section~\ref{sec:Method}. We present our observed TPCFs and their evolutionary changes in Section~\ref{sec:Results} along with details on the statistical tools and toy models we adopt to understand the features in the TPCF. Following this, in Section~\ref{sec:InferredProps}, we calculate some key physical quantities that highlight differences in the overall spatial distribution of clusters among the galaxies, and attempt to compare this with the properties of the host galaxy. Finally, we summarise our findings in Section~\ref{sec:Summary}.

\section{Data}
\label{sec:Data}
\subsection{Galaxy Sample}\label{sec:galaxy_sample}
In this study, we select twelve local ($<18$ Mpc) galaxies from the LEGUS survey of various morphological types, ranging from irregular dwarfs to grand design spirals. The twelve galaxies were picked from the larger sample of LEGUS galaxies for which cluster catalogues were available based on the conditions that - i) they contain sufficient number of star clusters to calculate TPCFs across a range of separations, and ii) were relatively face-on to prevent line-of-sight inclination effects. The galaxies and their average physical properties are listed in Table~\ref{tab:galaxyinfo}, and we provide more detail on the individual galaxies below.

The LEGUS sample consists of both archival and new imaging with either the Wide Field Camera 3 (WFC3) or the Advanced Camera for Surveys (ACS). The F275W (UV) and F336W (U) filters of each galaxy in this sample are WFC3 imaging. The three other bands are taken with either the ACS (archival) or WFC3 (new observations for the LEGUS Programme GO--13364 \citealt{Calzetti_2015}): ACS/WFC3 F435W (B), ACS/WFC3 F555W/F606W (V), and ACS/WFC3 F814W (I). In this study, we refer to the passbands by the conventional Johnson passband naming: UV, U, B, V, and I, where the V-band is adopted as the reference frame. LEGUS photometry is in the Vega magnitude system. The frames are aligned and rotated with North up. Some of the galaxies are observed with more than one pointing and combined into a single mosaic, whereas others are observed with a single pointing. Reduced science frames in all filters have been drizzled to a common scale resolution, corresponding to the native WFC3 pixel size ($0\farcs03962$/px). 

\subsubsection{NGC 0628}

NGC~0628 is a nearly face-on spiral galaxy (morphology SAc) located at a distance of~$\sim 9.8$~Mpc \citep{Anand_2021}. It was observed by the LEGUS survey with two pointings. This galaxy has been observed extensively by all recent major surveys of interstellar gas and dust in nearby galaxies, including THINGS, HERACLES, SINGS, KINGFISH, EMPIRE, and with ALMA \citep{Walter_2008,Leroy_2009,Kennicutt_2003,Kennicutt_2011,Bigiel_2016,Turner_2019}. \citet{Elmegreen_2006} investigated hierarchical star formation in this galaxy, and \citet{Grasha_2015} report a measurement of its TPCF. 

\subsubsection{NGC 1313}
NGC~1313 is a mildly inclined barred galaxy (morphology SBd) that may be interacting with a satellite, producing a loop of HI gas around the galaxy \citep{Peters_1994} and a recent increase of the SFR in the south-west arm \citep{Silva-Villa_2012}. Due to both its physical and morphological properties, including the presence of a bar and an irregular appearance, NGC~1313 has been compared to the Large Magellanic Cloud \citep{Vaucouleurs_1963}. \citet{Hannon_2019} and \citet{Messa_2021} have recently analysed the properties of NGC~1313's star clusters and \ion{H}{ii} regions, using H$\alpha$ narrow-band and Near-Infrared (NIR) Pa$\beta$ observations. LEGUS observed NGC~1313 with two distinct pointings, and we use a mosaic prepared from the two pointings for the analysis in this study. 

\subsubsection{NGC 1566}
NGC~1566, the brightest member of the Dorado group, is an almost face-on spiral galaxy with an intermediate-strength bar and open, knotty arms, a small bulge, and an outer pseudo-ring made from arms that wind anti-parallel to the bar ends \citep{Buta_2015}. \citet{Salo_2010} propose that the spiral arms are formed through bar-driven spiral density waves, and \citet{Shabani_2018} find evidence for an age gradient in the star clusters across the spiral arms consistent with the stationary density wave theory. \citet{Grasha_2017_Spatial} and  \citet{Gouliermis_2017} use LEGUS catalogues to study the hierarchical distribution of star clusters and young stellar populations, respectively. We caution that there is significant uncertainty in the distance to the galaxy, with published estimates varying from $5.5$ to $21.3$~Mpc \citep{Tully_1988,Mathewson_1992,Willick_1997,Theureau_2007,Sorce_2014,Tully_2013,Calzetti_2015,Sabbi_2018, Anand_2021}. Here we adopt the value obtained from the Kourkchi-Tully group catalogue \citep{Kourkchi_2017}, i.e., $17.7$~Mpc, which uses a distance to the Dorado galaxy group obtained through numerical modelling of its orbits. That said, the only effect of changing the adopted distance for our study is that it would shift the correlation functions we obtain along the linear distance axis, and the corresponding conversion from angular separation to linear separation. This galaxy is observed with a single pointing by LEGUS. 

\subsubsection{NGC 3344}
NGC~3344 is an isolated barred spiral galaxy (morphology SABbc) with two ring-like morphological features at $1$ and $7$~kpc, and a small bar within the inner ring \citep[see, for e.g.][]{Montenegro_2000}. This galaxy was included in the sample studied in \citet{Grasha_2017_Spatial}. \citet{Meidt_2009} analysed the spiral structure and dynamics in this galaxy using $\ion{H}{i}$ and $\mathrm{CO}$ gas. The LEGUS survey observed NGC~3344 with a single-pointing. 

\subsubsection{NGC 3627}
NGC~3627 is a strongly barred spiral galaxy \citep{Buta_2015} that has been studied in large atomic and molecular gas surveys \citep{Walter_2008,Leroy_2009,Kennicutt_2011} and exhibits strong burst signatures at the two interfaces of bar and arm in the north and south \citep{Kennicutt_2011}, making it a prime candidate for the study of bar-arm interactions \citep{Beuther_2017}. In addition, NGC~3627 appears to be interacting with the neighbouring galaxy NGC~3628 \citep[see, e.g.,][]{Soida_2001}, which is expected to be the cause of a perturbed morphology of its western arm, and a higher H$_2$/H$_{\rm I}$ mass ratio relative to other local star-forming galaxies \citep[e.g.,][]{Saintonge_2011}. LEGUS observed this galaxy in a single-pointing.

\subsubsection{NGC 3738}
NGC~3738 is an irregular dwarf galaxy (morphology Im) in the Messier~81 group classified as a blue compact dwarf (BCD). It is close to the Milky Way ($\sim 9.9$~Mpc), and has a relatively small size ($R_{25} \sim 4$~kpc). \citet{Hunter_2012} included this galaxy in the LITTLE THINGS HI survey, and reported that the HI component of NGC~3738 is morphologically and kinematically disturbed, possibly a result of an advanced merger or ram pressure stripping \citep{Ashley_2017}. In addition, \citet{Hunter_2018} found that the gas pressure, density and star formation rate in a localised region in the south-west part of the galaxy is much higher than the rest of the galaxy, with a higher fraction of younger clusters found there. LEGUS observes the entire extent of the optical galaxy in a single pointing. 

\subsubsection{NGC 4449}

NGC~4449 is an irregular barred starburst galaxy (morphology SBm) with ongoing and intense star formation distributed along a bar-like structure with two streams stemming from its ends. The gas component shows morphological features that may be caused by dynamical interactions with neighbouring galaxies \citep{Hunter_1998}. It has a rich population of young, intermediate, and old star clusters, making it the best sampled dwarf galaxy in our study, potentially due to a rich star formation history sculpted by earlier interactions and mergers \citep[see,for e.g.,][]{Cignoni_2019}. \citet{Whitmore_2020} include this galaxy in the recent H$\alpha$-LEGUS survey that add narrowband H$\alpha$ imaging to a subsample of LEGUS galaxies, allowing the production of new cluster catalogues with improved ages. However, since age accuracy is not a significant constraint for our analysis, we choose to use the original LEGUS catalogues for consistency with the remainder of the sample. 

\subsubsection{NGC 5194}
NGC~5194 (M51a or the Whirlpool galaxy) is a well-studied spiral galaxy (morphology SAbc) due to its large size, relative proximity and almost face-on inclination. This galaxy contains the largest number of clusters in our sample. Its grand-design morphology, high star formation rate, rich star formation history, and numerous star-forming complexes and star clusters make it a benchmark for nearby extragalactic surveys \citep[e.g., PAWS][]{Schinnerer_2013}. A number of authors have investigated  NGC~5194's TPCF \citep{Bastian_2005,Scheepmaker_2009} as well as cross-correlations between clusters and molecular clouds \citep{Grasha_2019}. In addition, \citet{Messa_2018a} and \citet{Messa_2018b}, as part of the LEGUS survey, study the age and mass distributions of the young star cluster population and their dependencies on the local environment within the galaxy. NGC~5194 is known to be interacting with its companion galaxy (NGC~5195) resulting in a marked spiral geometry along with a tidal tail due to the interaction, and the two galaxies together are referred to as the M51 system. We note that the LEGUS field-of-view was obtained through multiple pointings, and the catalogued star clusters cover both members of the system. However, we found that removing the contribution from NGC~5195 star clusters does not change the TPCF, since it contains a very small fraction of the overall catalogue, and hence we keep the overall catalogue for completeness. We also refer to the system simply as NGC~5194 since this is the major contributor to the observed TPCF, and doing so maintains consistency in the format of names for the galaxies in this study. 

\subsubsection{NGC 5253}
NGC~5253 is a nearby blue compact dwarf galaxy that hosts a very young central starburst, likely triggered by infalling material along the minor axis of the galaxy \citep{Meier_2002,Miura_2015,Turner_2015,Miura_2018}. This results in a dense, clumpy, central region, hosting a rich population of dense super star clusters (SSCs) with very high star formation efficiencies \citep{Turner_2004,Calzetti_2015_5253,Turner_2017,Smith_2020}. This galaxy has the lowest number of catalogued clusters in this study, with an age distribution that skews young. The radial extent of the star cluster population is entirely covered in the LEGUS field-of-view, observed with a single pointing.

\subsubsection{NGC 5457}
NGC~5457, commonly referred to as the Pinwheel Galaxy, is a relatively large, almost face-on ($i \sim 18 \degr$) SABcd-type spiral galaxy with a complicated arm structure and a highly asymmetric disk morphology suggestive of previous accretion or interaction  \citep{Waller_1997,VanderHulst_1998,Walter_2008}. It has 823 catalogued star clusters spread across the extent of its large disk. Due to its large angular size, LEGUS covers this galaxy with 5 different pointings: 1 in the central region, 3 in the north-west, and 1 in the south-east. In this study, we only use the star clusters in the region spanned by the available mosaic galaxy image available on the LEGUS public website \footnote{\url{https://archive.stsci.edu/prepds/legus/dataproducts-public.html}}, since our analysis method requires knowledge of the mosaic footprint (see below). This mosaic does not span the south-east regions of the galaxy, and hence we exclude these star clusters from our TPCF analysis. Overall, this results in a relatively lower completeness in the azimuthal and radial sampling of the young clusters as compared to the smaller galaxies in our sample. 

\subsubsection{NGC 6503}
NGC~6503 is a spiral galaxy classified as an SAcd type in \citet{deVaucouleurs_1991}. It has well-developed spiral arms and traces of a bar \citep{Buta_2015}. The galaxy has a patchy circum-nuclear appearance in the gas distribution, a morphology that carries over to the young star cluster distribution observed with LEGUS. \citet{Freeland_2010} interpreted this as an inner ring around the galactic bar. \citet{Gouliermis_2015} studied the hierarchical distribution of unbound stars with the LEGUS survey, and found that younger stars are organised in a distribution with a 2D fractal dimension of 1.7, whereas older stars display a homogeneous distribution, with a structure dispersion timescale of $\sim 60 \, \mathrm{Myr}$. NGC~6503 also has a significant line-of-sight inclination, for which we compensate by de-projecting star cluster positions before computing the TPCF. This galaxy is fully covered with a single pointing. 

\subsubsection{NGC 7793}
NGC~7793 is a flocculent spiral galaxy (morphology SAd). It is part of the Sculptor group, and is one of the closest galaxies in the LEGUS sample. It is characterised by diffuse, broken spiral arms with no bar, a very faint central bulge, and a relatively low star formation rate. \citet{Sacchi_2019} study the star formation history of this galaxy using LEGUS data, while \citet{Grasha_2017_Spatial} and \citet{Grasha_2017_Ages} study the spatial and temporal TPCF of its star clusters. In addition, \citet{Grasha_2018} study the connection between molecular clouds and young star clusters, and the timescales of their mutual association in this galaxy. LEGUS observed this galaxy with two pointings, one each in the eastern and western parts of the galaxy. 

\renewcommand{\arraystretch}{1.7}
\begin{table*}
\centering
\caption{Summarised physical quantities of the galaxies in this study.}
\label{tab:galaxyinfo}
\begin{threeparttable}
\begin{tabular}{l c c c c c c c c c c c}
\toprule
\multicolumn{1}{l}{Name(1)}& \multicolumn{1}{c}{Morph.(2)}& \multicolumn{1}{c}{$T$(3)} & \multicolumn{1}{c}{$i$(4)} & \multicolumn{1}{c}{P.A.(5)} & \multicolumn{1}{c}{$D$(6)} & \multicolumn{1}{c}{$\mathrm{SFR}_{UV}$(7)} & \multicolumn{1}{c}{$M_{*}$(8)} & \multicolumn{1}{c}{$R_{25}$(9)} & \multicolumn{1}{c}{$\Sigma_{\mathrm{SFR}}$(10)} & \multicolumn{1}{c}{$\Sigma_{\mathrm{HI}}$(11)} &  \multicolumn{1}{c}{$N_{\mathrm{cl}}$(12)}  \\ 
\multicolumn{1}{l}{}& \multicolumn{1}{c}{}& \multicolumn{1}{c}{} & \multicolumn{1}{c}{[deg]} & \multicolumn{1}{c}{[deg]} & \multicolumn{1}{c}{[Mpc]} & \multicolumn{1}{c}{$ \left[ M_{\sun} \, \mathrm{yr}^{-1} \right]$} & \multicolumn{1}{c}{$ \left[ M_\odot \right] $} & \multicolumn{1}{c}{[kpc]} & \multicolumn{1}{c}{$ \left[ M_{\sun} \, \mathrm{yr}^{-1} \,\mathrm{kpc}^{-2} \right]$} & \multicolumn{1}{c}{$ \left[ M_{\sun} \, \mathrm{pc}^{-2} \right]$} & \multicolumn{1}{c}{} \\ 
\midrule
NGC 0628 &SAc &$5.2 $&8.9 &20.7 &$9.8 $&$3.67 $&$1.1 \times 10^{10}$&$15.0 $&$4.4 \times 10^{-3}$&$15.6$&$1262 $\\
NGC 1313 &SBd  &$7.0 $&51.0 &14.0 &$4.3 $&$1.15 $&$2.6 \times 10^{9}$&$5.7 $&$1.1 \times 10^{-2}$&$20.4$&$741 $\\
NGC 1566 &SABbc &$4.0 $&29.6 &214.7 &$17.7 $&$5.67 $&$2.7 \times 10^{10}$&$21.4 $&$3.9 \times 10^{-3}$&$4.0$&$1573 $\\
NGC 3344 &SABbc &$4.0 $&25.0 &155.0 &$9.8 $&$0.86 $&$5.0 \times 10^{9}$&$10.1 $&$2.7 \times 10^{-3}$&$7.1$&$396 $\\
NGC 3627 &SABb &$3.1 $&57.3 &173.1 &$11.3 $&$4.89 $&$3.1 \times 10^{10}$&$13.6 $&$2.3 \times 10^{-2}$&$2.6$&$742 $\\
NGC 3738 &Im &$9.8 $&22.6 &156.0 &$5.1 $&$0.07 $&$2.4 \times 10^{8}$&$1.9 $&$6.4 \times 10^{-3}$&$13.7$&$228 $\\
NGC 4449 &IBm &$9.8 $&45.0 &64.0 &$4.0 $&$0.94 $&$1.1 \times 10^{9}$&$3.6 $&$2.3 \times 10^{-2}$&$51.1$&$607 $\\
NGC 5194 &SAbc &$4.0 $&22.0 &173.0 &$8.6 $&$6.88 $&$2.4 \times 10^{10}$&$13.9 $&$1.7 \times 10^{-2}$&$3.8$&$3043 $\\
NGC 5253 &Im &$11.0 $&42.0 &21.0 &$3.3 $&$0.10 $&$2.2 \times 10^{8}$&$2.4 $&$5.5 \times 10^{-3}$&$5.5$&$80 $\\
NGC 5457 &SABcd &$6.0 $&18.0 &39.0 &$6.7 $&$6.72 $&$1.9 \times 10^{10}$&$27.9 $&$4.4 \times 10^{-3}$&$7.8$&$823 $\\
NGC 6503 &SAcd &$5.8 $&75.1 &135.0 &$6.3 $&$0.32 $&$1.9 \times 10^{9}$&$6.4 $&$2.5 \times 10^{-3}$&$10.1$&$298 $\\
NGC 7793 &SAd &$7.4 $&55.0 &98.0 &$3.6 $&$0.52 $&$3.2 \times 10^{9}$&$4.9 $&$6.8 \times 10^{-3}$&$10.3$&$371 $\\
\bottomrule
\end{tabular}
\begin{tablenotes}
\small
\item \textbf{Notes}:(1),(2): Galaxy name and morphological class as listed in the NASA Extragalactic database (NED).
(3): RC3 morphological T-type as listed in Hyperleda.
(4): Inclination angle in degrees. References for adopted inclinations in order of the rows: \citet{Lang_2020,Koribalski_2018,Lang_2020,Meidt_2009,Lang_2020,Oh_2015,Hunter_1998,Colombo_2014,Koribalski_2018,Walter_2008,Greisen_2009,Koribalski_2018}.
(5): Position angle measured anti-clockwise from the celestial north. References for adopted angles identical to the inclination angles, except for NGC 3738 which adopts the value reported in \citet{Vaduvescu_2005}.
(6): Redshift-independent distances adopted from \citet{Anand_2021} for all galaxies except NGC 3344, NGC 3738, NGC 5253, and NGC 6503 for which we use the value reported in \citet{Sabbi_2018}.
(7): Galaxy integrated Far-UV calculated Star formation Rate adopted from \citet{Calzetti_2015}.
(8): Stellar mass adopted from \citet{Calzetti_2015}.
(9): Standard isophotal radius of the galaxy adopted from \citet{deVaucouleurs_1991} after applying the distances reported in Column 6.
(10): Star formation rate surface density obtained by - i) averaging $\mathrm{SFR}_{\mathrm{UV}}$ uniformly in a disk of radius $R_{25}$ if the entire radial extent of the star-forming gas is contained in the LEGUS field of view, ii) averaging the local dust-extinction corrected $\mathrm{SFR}_{\mathrm{UV}}$ values only in the LEGUS field-of-view, if not. Values in this case are obtained from Adamo et al. (2021, in prep). 
(11): HI gas surface density obtained by averaging the total HI mass in the disk $\mathrm{M}_{\mathrm{HI}}$ uniformly in a disk of radius $R_{25}$. $\mathrm{M}_{\mathrm{HI}}$ values are adopted from \citet{Calzetti_2015}.
(12): Total number of identified star clusters in the LEGUS catalog that we use.
\end{tablenotes}
\end{threeparttable}
\end{table*}

\subsection{LEGUS Star Cluster Catalogues}
\label{sec:catalogs}
A detailed description of the standard data reduction of the LEGUS sample can be found in \citet{Calzetti_2015} and in-depth descriptions of the cluster extraction, classification, photometry, and SED fitting procedure are detailed in \citet{Adamo_2017}. The procedure to obtain the catalogues for the dwarf galaxies are given in \citet{Cook_2019}. We refer the reader to these papers and provide a brief description of the LEGUS cluster catalogues here. 

\subsubsection{Automated Cluster Catalogue Procedure} 
Catalogue construction in LEGUS is a multi-step process that begins with an initial automated extraction of cluster candidates identified using Source Extractor \citep[SEXTRACTOR;][]{Bertin_1996} from the white-light images produced with the five standard LEGUS bands \citep{Calzetti_2015}. The SEXTRACTOR parameters are optimized to extract sources with at least a 3$\sigma$ detection in a minimum of five contiguous pixels. 

The automatic catalogues for each galaxy includes sources that satisfy the two following conditions: (1) the V-band concentration index (CI $\equiv$ magnitude difference of a source in an aperture of 1 pixel compared to an aperture of 3 pixels) must be greater than the stellar CI peak value; and (2) the source must be detected in at least two contiguous filters (the reference V band and either B or I band) with a photometric error $\sigma_{\lambda} \leq 0.35$. These conditions minimize stellar contamination and yield cluster candidates with a signal-to-noise greater than 3, which allows for reliable constraints on the derived cluster properties of age and mass. This procedure produces our automated cluster catalogue that is complete for clusters down to 1~pc in size for galaxies at distances up to 10~Mpc \citep{Adamo_2017}. This size is well below the peak of the size distribution of star clusters of $\sim$3~pc \citep{Ryon_2017}.

\subsubsection{Photometry}
The next step in catalogue construction is photometry. The analysis pipeline measures the luminosity of each cluster using a science aperture of radius 4--6 pixels depending on the distance to the galaxy, with sky corrections computed using a sky annulus at 7~pixels with a width of 1~pixel. The pipeline then applies an average aperture correction, which it estimates from the difference between the luminosity within the science aperture and that within a 20-pixel aperture with a 1-pixel sky annulus for a control sample of isolated clusters. The pipeline applies this aperture correction independently in each filter. The photometry reported in the catalogue is in the Vega magnitude system and is also corrected for foreground Galactic extinction \citep{Schlafly_2011}. 

\subsubsection{Estimation of cluster masses and ages}

To ensure that we can derive reliable estimates of cluster physical properties (age, mass, and extinction), the next analysis step is to remove from the catalogue any clusters that lack a 3$\sigma$ detection in at least four of the five photometric bands. The pipeline then estimates the masses and ages of the remaining clusters by fitting the observed SED using Yggdrasil deterministic stellar population models \citep{Zackrisson_2011} using a $\chi^2$ fitting approach that includes uncertainty estimates \citep{Adamo_2010, Adamo_2012}. The uncertainties derived in the physical parameters for the final LEGUS star clusters are on average 0.1~dex \citep{Adamo_2017}. The Yggdrasil models are based on Starburst99 \citep{Leitherer_1999} stellar population spectra coupled with nebular emission computed using Cloudy \citep{Ferland_1998, Ferland_2013}. \citet{Adamo_2010} adopt a \citet{Kroupa_2001} IMF in the range 0.1--120~M$_\odot$ \citep[see, however, ][for a generalisation to a variable IMF]{Ashworth_2017}. We also adopt the \citet{Adamo_2017} catalogs that use the Padova stellar isochrones that include thermally pulsating asymptotic giant branch stars \citep{Vazquez_2005, Girardi_2000} and a starburst attenuation curve \citep{Calzetti_2000} with the assumption that stars and gas undergo the same amount of reddening. \citet{Adamo_2017} computes the nebular emission assuming a hydrogen number density n$_\textrm{H}=10^2$~cm$^{-2}$, a covering factor $c = 0.5$ (i.e., 50\% of the Lyman continuum photons that are produced by the central source driving nebular emission from within the LEGUS aperture), and a gas filling factor $f$ of 0.01, typical of H\,\textsc{ii} regions \citep{Croxall_2016}. 

\subsubsection{Visual classification}

\begin{figure*}
    \centering
    \includegraphics[width=\textwidth]{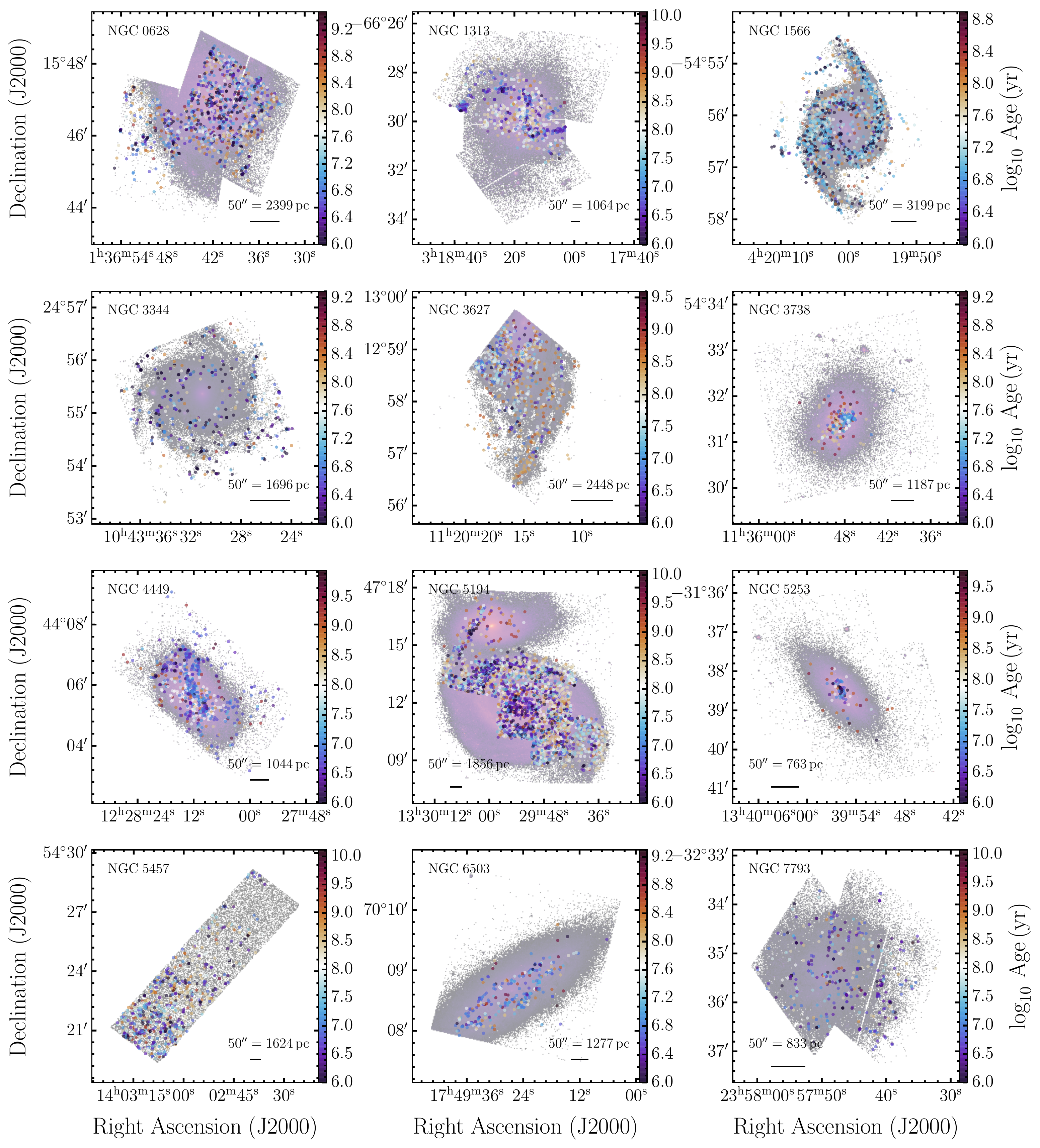}
    \caption{Plane-of-sky positions of star clusters catalogued with LEGUS, coloured by the age of the cluster, overplotted on the \textit{HST} image of the galaxy. A scale bar denoting an angular size of $50 \arcsec$, with the corresponding linear separation obtained by using the distance to the galaxy reported in Table~\ref{tab:galaxyinfo}, is provided for each galaxy. Orientation of the image is celestial north up and east to the left of the image.}
    \label{fig:cluster_positions}
\end{figure*}

The final step in the analysis is visual classification. Members of the LEGUS team visually examine cluster candidates in the automated cluster catalogue if they satisfy the following two criteria: (1) detection in a minimum of four bands (VBI and U and/or UV) with a S/N above 3 sigma and (2) brighter than $-$6 mag in the V-band \citep{Grasha_2015, Adamo_2017}. The cluster catalogue for NGC~5194 is obtained with a combination of visual and Machine Learning (ML) procedures for the final cluster classifications  \citep{Grasha_2019}. The human or machine classifiers assign each cluster to one of four morphological classes: Class~1 contains compact, symmetric, and centrally concentrated clusters. Class~2 includes compact clusters with asymmetry. Class~3 are compact associations that show multiple-peaked profiles on top of an underlying diffuse emission. Class~4 is the label given to non-cluster contaminants that remain in the catalogue after all selection criteria. These are usually bad pixels, foreground stars, or background galaxies. Star cluster candidates that are not visually inspected are labelled as Class~0 in the final catalogues. In this study, we consider all cluster candidates (class~1, 2, and 3) for our results and analysis and do not separate by cluster classification type. We note that some of the candidates we chose to represent as "star clusters", particularly class~3 ones, may not be gravitationally bound and could disperse or dissolve in relatively short timescales ($\sim 10 \mathrm{Myr}$). However, since we are interested in the hierarchy of star formation and the spatial distribution of star formation, including these unbound associations in our definition of star clusters is warranted. The total number of such star clusters in each galaxy is listed in Table~\ref{tab:galaxyinfo}. The positions of the identified star clusters in the plane-of-sky is shown in Figure~\ref{fig:cluster_positions}, overplotted on the \textit{HST} image of the galaxy.   

The magnitude limit of $-6$ in V that we set for visual inspection corresponds to a 1000~M$_\odot$, 6~Myr star cluster with colour excess $E(B-V)=0.25$ \citep{Calzetti_2015}. Thus our catalogues are incomplete at lower masses. However, this limit corresponds approximately to the completeness limit of the underlying photometric catalogue. \citet{Adamo_2017} carry out artificial cluster tests for NGC~628 (distance of 10~Mpc), and find that LEGUS produces a complete cluster sample down to a cluster mass of 5000~M$_\odot$ for cluster ages $<200$ Myr. Since our clustering results are driven by much younger clusters ($<$10~Myr), where we are complete down to even lower masses, incompleteness will have minimal impact on the results and analysis.

The final catalogues of which we make use in this work are publicly available online\footnote{\url{https://archive.stsci.edu/prepds/legus/dataproducts-public.html}} on the Mikulski Archive for Space Telescopes (MAST) for all galaxies except NGC 3627 and NGC 5457. The catalogue for these galaxies will be published in a forthcoming paper (Linden et al., in prep).

\section{Methods}
\label{sec:Method}

\subsection{Deprojection}
\label{sec:inclination_correction}

It is important for our analysis of the spatial distribution of clusters to de-project the cluster positions from the plane of the sky to the plane of the galaxy. This is necessary, especially for higher inclination angles, as the inclination modifies the true spatial separations between the clusters in the plane of the galaxy. The first step in our analysis is therefore to deproject cluster positions.

To perform this correction, we assume that each galaxy can be described with an axisymmmetric flat rotating elliptical disk. We then correct the position of each star cluster in a two step process. First, we rotate the intrinsic positions of the clusters by an angle $\phi$ in the clockwise direction about the centre of the galaxy, where $\phi$ is the position angle measured anti-clockwise from the celestial north, to align the major axis of the galaxy in the North-South direction. For a cluster with RA and DEC positions $x$ and $y$ relative to the centre of the galaxy, with $x$ increasing along RA (towards the left of Figure~\ref{fig:cluster_positions}) and $y$ along DEC (towards north of Figure~\ref{fig:cluster_positions}), we compute the new positions as
\begin{align}
    x' = x \cos(\phi) + y \sin(\phi) \\
    y' =  y \cos(\phi) - x \sin(\phi),
\end{align}
where $x'$ and $y'$ are the position-angle corrected RA and DEC of the cluster. Our second step is to correct for the line-of-sight inclination angle $i$ by dividing $x'$ by $\cos i$ while leaving the $y$ position unchanged. Thus the final positions of all clusters are $x_i = x'/\cos i$ and $y_i = y'$. The values of $\phi$ and $i$ that we use for each galaxy are provided in Table~\ref{tab:galaxyinfo}. The RA and DEC of the galaxy centres are taken to be the reported sky coordinates for the galaxies in the NASA/IPAC Extragalactic Database (NED) \footnote{\url{http://ned.ipac.caltech.edu}}. We use these deprojected cluster positions $(x_i, y_i)$ for all calculations of the correlation function in this paper.

\subsection{Angular TPCF}
To investigate the hierarchical distribution of young star clusters in galaxies, we use the angular Two-Point Correlation Function (TPCF) $1+\omega(\theta)$, where $\theta$ is the angular separation between a pair of clusters in the plane-of-sky. We refer to this quantity as the TPCF or the correlation function interchangeably for the remainder of the paper. The physical meaning of the TPCF is that, if one examines an annulus of radius $\theta$ and infinitesimal width $d\theta$ centred on one cluster, $1+\omega(\theta)$ is the ratio of the probability of finding another cluster within this annulus to the probability of finding one in an identical annulus that is placed at a random position, rather than centred on a known cluster. Mathematically, we express this by writing the conditional probability $dP(\theta)$ that a pair of clusters in a region is separated by an angle $\theta$ as,   
\begin{equation}
    dP(\theta) = \langle N \rangle^2 \left( 1+ \omega \left(\theta \right) \right) d\Omega_1 d\Omega_2,
\end{equation}
where $\langle N \rangle$ is the average surface density of clusters in the region per steradian, $d\Omega_1$ and $d\Omega_2$ are infinitesimal solid angle elements around clusters 1 and 2 respectively, and $1+\omega(\theta)$ has the typical form of a correlation function, 
\begin{equation}
    1+\omega(\theta) = \frac{\left< N(\theta_1) N(\theta_1+\theta) \right>}{\langle N \rangle^2},
\end{equation}
where $N(\theta_1)$ and $N(\theta_1+\theta)$ are the local surface densities around two regions separated by angle $\theta$, and the averaging is done over all such pairs of regions. From the definitions of the quantities above, we can interpret $\omega(\theta)$ as the quantity that represents the excess probability above a purely random Poisson distribution of finding a pair of points separated by an angle $\theta$. Indeed, for a purely random distribution, $\omega(\theta) = 0$, and by corollary, the correlation function $1+\omega(\theta) = 1$. For a clustered distribution $\omega(\theta)>0$, and for a scale-free clustered distribution such as a fractal, the correlation function $1+\omega(\theta)$ is a pure power law with a negative slope, up to the scale at which the distribution remains a fractal \citep{Calzetti_1988}. 

A number of authors have proposed estimators to calculate $\omega(\theta)$ for a given observed distribution of pointlike objects \citep{Peebles_1974,Peebles_Hauser_1974,Sharp_1979,Shanks_1980,Hewett_1982,Hamilton_1992,Landy_1993}. We use the one proposed by \citet{Landy_1993} as it attempts to correct for effects near the edge of the field-of-view, and provides estimates whose errors are largely Poisson distributed. It uses a combination of the data sample and a random sample that populates the field of view of the data. The Landy-Szalay estimator (LS, hereafter) is calculated as 
\begin{equation}
\label{eq:omega_eqn}
\omega_{\mathrm{LS}}(\theta)=\frac{\mathrm{DD}(\theta)-2 \mathrm{DR}(\theta)+\operatorname{RR}(\theta)}{\operatorname{RR}(\theta)}
\end{equation}
where DD is the number of data-data pairs, DR is the number of cross-correlated data-random pairs, and RR the number of random-random pairs, counting all pairs in the range of separations $\theta \pm d\theta$, where $d\theta$ is the adopted width of the discrete separation bin for which the TPCF is calculated. It is typically desirable to have a large enough random sample to cross correlate with, so as not to introduce any additional Poisson error in the estimator. To accommodate this, we normalise the counted pairs at a separation $\theta$ to the total number of possible pairs in the data, random, and cross-correlated distributions. This leads to the following modifications to the definition of DD, DR and RR
\begin{equation}
\begin{array}{l}
\mathrm{DD}(\theta)=\frac{P_{\mathrm{DD}}(\theta)}{N_D(N_D-1)} \\
\mathrm{DR}(\theta)=\frac{P_{\mathrm{DR}}(\theta)}{N_D N_{R}} \\
\operatorname{RR}(\theta)=\frac{P_{\mathrm{RR}}(\theta)}{N_{R}\left(N_{R}-1\right)},
\end{array}
\end{equation}
where $N_D$ and $N_R$ denote the total number of data and random points in the distribution. We ensure that the sky coverage and geometry of the random sample is as identical as possible to that of the data, to allow accurate TPCF computation, especially close to the edge of the field of view. This is done by preparing a Poisson distribution of points occupying the HST footprint of the observed galaxy, then masking out any unsampled regions (if any) that happen to fall on the chip gaps of the ACS instrument. The footprint of the galaxy is prepared from the observed $V$-~band image using the \verb|FootprintFinder|~\footnote{\url{http://hla.stsci.edu/Footprintfinder/FootprintFinder.html}} tool that is publicly available, and adjusted to account for the deprojection of the galaxy. We do not, however, mask potentially dust-extincted regions in the galaxy such as dust lanes, since \citet{Grasha_2015} found that doing so does not affect the resulting TPCF. 

To compute the value of the TPCF we use our own modified version of the TPCF functionality offered by the python \verb|astroML|~\footnote{\url{https://www.astroml.org/index.html}} module \citep{astroML}, which uses the \verb|scikit-learn|~\footnote{\url{https://scikit-learn.org/stable/index.html}} library as a backend for fast computations of pairs at given separations \citep{scikit-learn}. We use a bootstrap method \citep{Efron1994} with 100 bootstrap samples to estimate the value of and error bars on $\omega$ in each bin. The code we used to perform the analysis is publicly available on \verb|Github|~\footnote{\url{https://github.com/shm-1996/legus-tpcf}}. 

\subsection{Edge Effects}
\label{sec:edgeeffects}
At angular separations that approach the size of the telescope field-of-view, the TPCF is prone to bias due to edge effects. The \citet{Landy_1993} estimator attempts to compensate for this by smoothing the steep fall in the TPCF expected near the edge, as the number of pairs in the data goes to zero near the boundary. However, this correction is not perfect, and hence separations where it is significant should be interpreted with caution. To obtain an estimate for the scale where this correction starts to matter, we perform numerical simulations of toy 2D fractal distributions truncated by square fields of view of different sizes, and estimate the size $l_{\rm edge}$ at which edge effects become significant. We provide details of the procedure we use to estimate this scale in Appendix~\ref{sec:appendix_edge}. We find that $l_{\mathrm{edge}} \sim R_{\mathrm{max}}/5$, where $R_{\mathrm{max}}$ is the size of the field-of-view, and the value of the TPCF for scales beyond $l_{\mathrm{edge}}$ has a significant contribution from smoothing by the estimator we use. Thus, in all our plots of TPCFs, we use grey shading to indicate separations $\theta > \theta_{\mathrm{max}}/5$, where $\theta_{\mathrm{max}}$ is the angular extent of the deprojected \textit{HST} field-of-view footprint. In many cases the deprojected field-of-view is not a square, in which case we use the longest side for $\theta_{\mathrm{max}}$. The separation $\theta_{\mathrm{max}}/5$ is used to caution the reader about scales where edge effects might play a role and the TPCF should be interpreted with caution.

\section{Results}
\label{sec:Results}

\subsection{Observed TPCFs of Galaxies}
\label{sec:tpcf_galaxies}

We compute the angular TPCF $1+\omega({\theta})$ for star clusters in the galaxies of our sample through the method outlined in Sec.~\ref{sec:Method}, using $20$ angular separation bins that are logarithmically spaced in the range \mbox{$10$--$5000$}~pc. We use the same bins (in linear rather than angular units) across all the galaxies to allow for consistent comparison of the TPCFs. Our choice to use 20 bins represents a compromise between resolving finer features of the TPCF and avoiding excessive shot noise. We mask bins for which the value of $1+\omega \leq 0$, which we sometimes encounter at large separations when the number of pairs in the data are very low. We also do not show the TPCFs for bins where the median value of $\omega$ is lower than the bootstrap-calculated error in $\omega$, a condition that arises occasionally for narrow, low separation bins where the number of pairs is small. 

\begin{figure*}
    \centering
    \includegraphics[width=0.98\textwidth]{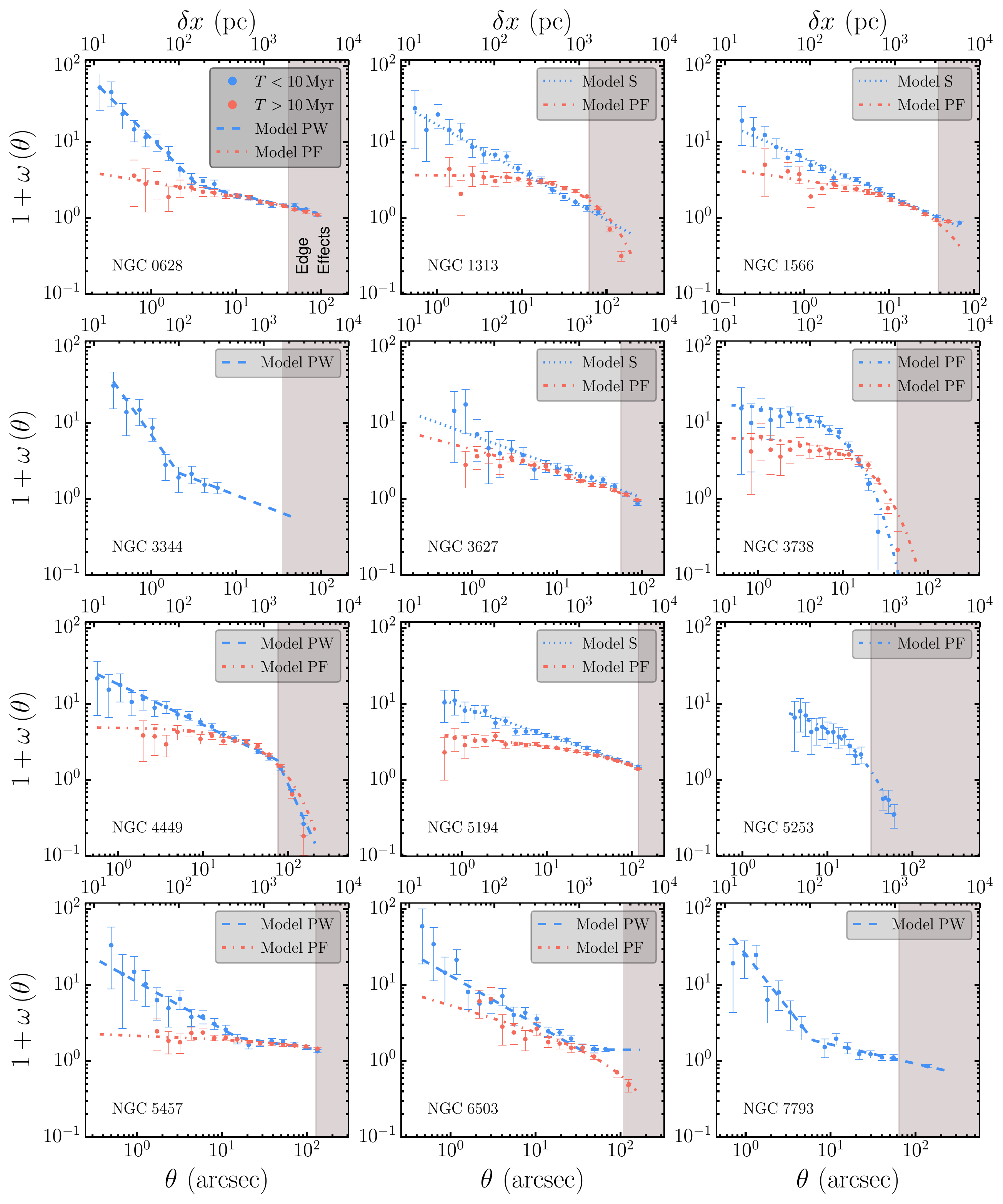}
    \caption{The TPCF $1+\omega(\theta)$ of star clusters with ages $T \la 10 \, \mathrm{Myr}$ (blue) and $T>10\, \mathrm{Myr}$ (red) for each galaxy, with the best fit functional form model for both age groups overplotted (dashed lines) in their corresponding colours. The functional models that are fitted are: Single power law (Model~S), PieceWise power law (Model~PW), and Power law with an exponential Fall-off (Model~PF), and are represented as dotted, dashed, and dot-dash line styles, respectively. The functional form among these that fits the TPCF best for both age groups is reported in the legend, with the best-fit parameters and superior model reported in Table~\ref{tab:AIC}, and the detailed fitting procedure outlined in Section~\ref{sec:fitting}. The bottom $x-$axis denotes $\theta$, the angular separation in arcsec, and the top axis denotes $\delta x$, the corresponding linear separation in parsec (pc) using the distance to the galaxy reported in Table~\ref{tab:galaxyinfo}. Grey shaded regions denote the estimated range of separations where edge effects could play a role in the TPCF (see Section~\ref{sec:edgeeffects}). }
    \label{fig:age_comparison}
\end{figure*}

In order to isolate evolutionary effects, for the bulk of this paper we will consider only TPCFs for clusters separated by age. Separating clusters by age is important because there are also physical differences between the galaxies that lead to perceptible variations in their TPCFs, even within the same age group. Computing the TPCFs of clusters of all ages as a single distribution tends to mix the two causes of variation, making it impossible to disentangle variations in galaxies spatial structure from variations in their star formation histories. However, for reader convenience we do present the TPCF for the combined cluster sample without binning by age in Appendix \ref{app:TPCF_all}. For the remainder of our analysis, we choose an age of 10~Myr to separate young ($T \la 10\, \mathrm{Myr}$) and old ($T>10\, \mathrm{Myr}$) clusters. This choice is motivated by recent evidence that star-forming giant molecular clouds (GMCs) are dispersed by feedback from massive stars in \mbox{$1$--$5 \, \mathrm{Myr}$} \citep[see,][and references therein]{Chevance_2020}, where we have chosen a conservatively higher value as some galaxies have a low number of clusters at ages $T<5 \, \mathrm{Myr}$, and our age estimates carry some uncertainty. Thus, our division of young and old clusters corresponds roughly to those that are probably still associated with their natal molecular cloud, and those that are not, respectively. While in principle this timescale would be different in each galaxy, we choose a consistent value for simplicity. In Figure~\ref{fig:age_comparison} we show the resulting TPCFs for young and old clusters in the galaxies. We note that NGC~3344, NGC~5253 and NGC~7793 have very few clusters with ages $T>10 \, \mathrm{Myr}$, and thus their TPCFs are extremely noisy even for the bins that have non-zero correlation. For this reason we omit the TPCF for older clusters in these galaxies from the figure.

We find that, in general, there are significant variations between the TPCF of young and old clusters for a given galaxy. For instance, younger clusters seem to have relatively higher values of correlation and seem to show, at least qualitatively, power-law behaviour (straight line in a log-log plot) as a function of separation $\theta$ over a range of scales. Older clusters, on the other hand, seem to show TPCFs that are relatively flat at small separations and show some form of smooth fall-off at large separations, suggesting some sort of evolutionary effect. However, there is also significant variation in the TPCFs for a given age group among the galaxies, especially for the young clusters group. While some galaxies show power-law behaviour in their young cluster TPCFs over the entire range of scales where we measure it, others seem to show a steep power law at small scales, followed by a shallow/flat power law at larger scales, and a relatively sharp break between the two regimes. In addition, there are some galaxies, specifically the dwarfs, that seem to show power-law behaviour at small separations, followed by a smooth fall-off at larger separations. These differences, we expect, should be due to physical differences in the underlying star cluster distribution, which is presumably set by the host galaxy. We intend to understand both forms of differences, evolutionary and physical, in the sections below. First, in Section~\ref{sec:Features}, we formally characterise the qualitative features we described above in the observed TPCFs, by fitting functional forms that reproduce these features, and using model comparison to choose the form that best describes the TPCF. We then attempt to infer the underlying distributions that might give rise to the observed TPCFs through the use of physically motivated toy model distributions. This allows us to infer the changes in the underlying star cluster distribution with host galaxy and with age. Following that, in Section~\ref{sec:tpcf_agecomparison}, we probe evolutionary effects in further detail for a single galaxy, by adding more age groups, and measuring the TPCFs for clusters that fall in them, in order to obtain finer time resolution.

\subsection{Quantitative Analysis of TPCFs}
\label{sec:Features}
In this section, we attempt to explain the observed features in the TPCFs outlined in Section~\ref{sec:tpcf_galaxies} by characterising them quantitatively (Section~\ref{sec:fitting}), and comparing with physically motivated toy model distributions to infer the underlying star cluster spatial distributions. The full description of the parameters of the toy models and how they influence the TPCF of the distribution is provided in Appendix~\ref{sec:appendix_toymodels}. Here, we just briefly describe and motivate the toy models for each identified feature, and infer the properties of the spatial distribution of the clusters from the comparison between the toy models and our measured TPCFs.

\subsubsection{Classifying TPCFs}
\label{sec:fitting}

The first step to understanding the TPCFs is to classify them by morphology of their features. We do this by defining three functional forms/models for the TPCF that represent qualitatively the three features described above, attempting to fit them to the observed TPCFs, and performing a statistical comparison of the fits to identify which functional model describes the data best. The three models can be described qualitatively as,
\begin{enumerate}
    \item Model S: A single power law with a fixed slope.  
    \item Model PW: A piecewise power law consisting of two fixed-slope segments separated by a transition/break point.  
    \item Model PF: A power law with an exponential cutoff to represent a power law that falls off smoothly at larger scales. 
\end{enumerate}
Quantitatively, we define model S as
\begin{equation}
\label{eq:ModelS}
    F_\mathrm{S} (\theta) = A_1 \theta^{\alpha_1},
\end{equation}
where $A_1$ is the amplitude and $\alpha_1$ the power-law slope. Model~PW is given by
\begin{equation}
\label{eq:ModelPW}
\begin{array}{l}
F_\mathrm{PW} (\theta) 
\quad=\left\{\begin{array}{ll}
A_1 \theta^{\alpha_1} & : \theta<\beta \\
A_2 \theta^{\alpha_2} & :\theta>\beta
\end{array}\right.
,
\end{array}
\end{equation}
where $\alpha_1$ and $\alpha_2$ are the two power-law slopes, $\beta$ the transition point, and $A_1$ and $A_2$ the amplitudes, which are related by $A_2 = A_1\beta^{\alpha_1-\alpha_2}$ to ensure continuity at $\theta=\beta$. Lastly, Model~PF is given by
\begin{equation}
\label{eq:ModelPF}
    F_\mathrm{PF} (\theta) =  A_1 \theta^{\alpha_1} \exp \left(- \frac{\theta}{\theta_c} \right),
\end{equation}
where the slope is as defined for the single power-law case, and $\theta_c$, is the scale above which the TPCF falls off. 

We fit all three model TPCFs to the observed correlation function $1+\omega(\theta)$ of every galaxy in our sample using a Markov-Chain Monte Carlo (MCMC) method \citep{Mackay2003}. The parameter vectors to fit for are $\lambda_\mathrm{S} = (A_1,\alpha_1)$, $\lambda_\mathrm{PW} = (A_1,\alpha_1,\alpha_2)$, and $\lambda_\mathrm{PF} = (A_1,\alpha_1,\theta_c)$ for Model~S, PW and PF, respectively, and the likelihood function is given by
\begin{equation}
\label{eq:likelihood}
    \ln \mathcal{L} = -\frac{1}{2} \sum \left( \frac{D(\theta) - M(\theta|\lambda) }{\sigma_D(\theta)} \right)^2,
\end{equation}
where $D(\theta)$ is the observed value of $1+\omega(\theta)$ at separation $\theta$, $M(\theta|\lambda)$ the corresponding model value at this separation for the parameter vector $\theta$ and $\sigma_D(\theta)$ the error in the observed TPCF value. The priors we use on the parameters are: $A_1>0$, $-5 < \alpha_1 \leq 0$, $ -5 < \alpha_2 \leq 0$, $\theta_{\mathrm{min}} < \beta <\theta_{\mathrm{max}}$, and $\theta_{\mathrm{min}}<\theta_c \leq 5\theta_{\mathrm{max}}$, where $\theta_{\mathrm{min}}$ and $\theta_{\mathrm{max}}$ are the minimum and maximum separations over which we compute the TPCF. We use the python package \verb|EMCEE| \citep{Emcee_2013} to perform the calculation, using 300~walkers, with a total of 5000~steps, discarding the first 200~steps as burn-in. We verified that the MCMC had reasonably converged in such a case through visual inspection of the MCMC chain.

To determine which model among the three described above is the best description for a given galaxy and age group, we compute the Akaike Information Criterion (AIC) \citep{Akaike_1974} for each fitted model. The AIC is an estimator for the relative quality of statistical models, given a set of data, which compares goodness-of-fit along with a factor that penalises for a higher number of parameters, preventing over-fitting a model to data. Given a set of candidate models, with their respective AIC values, the preferred model is the one with the minimum AIC value. In this study, we use the so-called corrected AIC \citep{Hurvich_1989}, which adds a correction term to the traditional AIC value, making it suitable for small sample sizes, and converges to the traditional AIC for an infinite sample size \citep{Burnham_2004}. It is given by 
\begin{equation}
\label{eq:AIC_eqn}
    \mathrm{AIC} = 2N_\lambda - 2 \ln(\mathcal{L}_{\mathrm{max}}) + \frac{2N_\lambda(N_\lambda+1)}{N_\theta - N_\lambda -1},
\end{equation}
where $N_\lambda$ is the number of parameters for a model, and $N_\theta$ is the number of angular separation bins for which the correlation function is calculated. In Table~\ref{tab:AIC}, we report for each galaxy and age group (i.e., young and old) the AIC values for all three models, the model with the minimum (best) AIC, and the resulting best-fit parameters for the superior model. In Figure~\ref{fig:age_comparison}, we overplot this best-fit model for both young and old clusters on their corresponding measured TPCFs with different line styles, and list the best-fit model name in the legend for each galaxy and age group. 

We find that the best-fit model for the older clusters in a galaxy is typically Model~PF, and this is more or less consistent across the sample. On the other hand, the best-fit model for younger clusters varies. Among the spirals, the TPCFs of NGC~1313, NGC~1566, NGC~3627 and NGC~5194 are best fit by Model~S, whereas NGC~0628, NGC~3344, NGC~5457, NGC~6503, and NGC~7793 prefer Model~PW. For the dwarf galaxies, NGC~3738 and NGC~5253 prefer Model~PF for both young and old clusters, and NGC~4449 is best fit by Model~PW and Model~PF for young and old clusters, respectively. To understand these differences and their physical implications it is important to first identify what sort of underlying star cluster distribution gives rise to the three models. We perform this exercise for all three fit models by comparing with physically motivated toy model distributions below. 

\renewcommand{\arraystretch}{1.7}
\begin{table*}
\caption{MCMC best-fit parameters and associated Akaike Information Criterion (AIC) values obtained by fitting the three models described in Section~\ref{sec:fitting}, to the observed TPCF of young ($\la 10 \, \mathrm{Myr}$) and old ($> 10 \, \mathrm{Myr}$) clusters.}
\centering
\label{tab:AIC}
\begin{threeparttable}
\begin{tabular}{l c c c c c c c c c}
\toprule
\multicolumn{1}{l}{Galaxy} & \multicolumn{1}{l}{Age Group} & \multicolumn{1}{c}{$\mathrm{AIC}_{\mathrm{S}}$} & \multicolumn{1}{c}{$\mathrm{AIC}_{\mathrm{PW}}$} & \multicolumn{1}{c}{$\mathrm{AIC}_{\mathrm{PF}}$} &\multicolumn{1}{c}{Best Model} & \multicolumn{1}{c}{$\alpha_1$} & \multicolumn{1}{c}{$\alpha_2$} & \multicolumn{1}{c}{$\beta$ ($\arcsec$)} & \multicolumn{1}{c}{$\theta_c$  ($\arcsec$)}\\ 
\midrule
\multirow{2}{*}{NGC 0628} &
Young &$135 $&$31.0 $ &$170 $ &PW &$-1.1^{+0.14}_{-0.14} $&$-0.3^{+0.02}_{-0.02} $&$3.9^{+1.4}_{-1.2} $&$-$\\
& Old &$15 $&$17 $ &$12 $ &PF &$-0.2^{+0.02}_{-0.02} $&$-$ & $-$ & $380^{+50}_{-71} $\\
\hline
\multirow{2}{*}{NGC 1313} &
Young &$12 $&$16 $ &$19 $ &S &$-0.6^{+0.03}_{-0.03}$ & $-$ & $-$ & $-$\\
& Old &$212$&$19 $ &$16 $ &PF &$-0.0^{+0.02}_{-0.04} $&$-$ & $-$ & $88^{+11}_{-7} $\\
\hline
\multirow{2}{*}{NGC 1566} &
Young &$18 $&$22 $ &$23 $ &S &$-0.5^{+0.02}_{-0.02}$ & $-$ & $-$ & $-$\\
& Old &$29$&$17 $ &$14 $ &PF &$-0.2^{+0.07}_{-0.06} $&$-$ & $-$ & $62^{+32}_{-18} $\\
\hline
\multirow{2}{*}{NGC 3344} &
Young &$22$&$20 $ &$>10^{3}$&PW &$-1.4^{+0.32}_{-0.37} $&$-0.5^{+0.27}_{-0.30} $&$2.2^{+2.0}_{-1.3} $&$-$\\
& Old &$-$ & $-$ & $-$ & $-$ & $-$ & $-$ & $-$ \\
\hline
\multirow{2}{*}{NGC 3627} &
Young &$15 $&$19 $ &$17 $ &S &$-0.4^{+0.04}_{-0.04}$ & $-$ & $-$ & $-$\\
& Old &$19$&$20 $ &$17 $ &PF &$-0.3^{+0.03}_{-0.02} $&$-$ & $-$ & $360^{+58}_{-73} $\\
\hline
\multirow{2}{*}{NGC 3738} &
Young &$81 $&$76 $ &$15 $ &PF &$-0.1^{+0.04}_{-0.09} $&$-$ & $-$ & $9.5^{+1.4}_{-1.0} $\\
& Old &$67$&$>10^{3}$&$28 $ &PF &$-0.0^{+0.02}_{-0.05} $&$-$ & $-$ & $21^{+3}_{-2} $\\
\hline
\multirow{2}{*}{NGC 4449} &
Young &$86 $&$17 $ &$24 $ &PW &$-0.5^{+0.03}_{-0.03} $&$-2.4^{+0.3}_{-0.4} $&$74^{+1}_{-1} $&$-$\\
& Old &$128$&$>10^{3}$&$38 $ &PF &$-0.0^{+0.01}_{-0.02} $&$-$ & $-$ & $69^{+5}_{-4} $\\
\hline
\multirow{2}{*}{NGC 5194} &
Young &$16 $&$18 $ &$18 $ &S &$-0.4^{+0.01}_{-0.01}$ & $-$ & $-$ & $-$\\
& Old &$56$&$21 $ &$13 $ &PF &$-0.1^{+0.02}_{-0.02} $&$-$ & $-$ & $295^{+51}_{-38} $\\
\hline
\multirow{2}{*}{NGC 5253} &
Young &$18 $&$16 $ &$11 $ &PF &$-0.6^{+0.34}_{-0.37} $&$-$ & $-$ & $36^{+68}_{-13} $\\
& Old &$-$ & $-$ & $-$ & $-$ & $-$ & $-$ & $-$ \\
\hline
\multirow{2}{*}{NGC 5457} &
Young &$31 $&$19 $ &$41 $ &PW &$-0.6^{+0.11}_{-0.17} $&$-0.2^{+0.05}_{-0.05} $&$14^{+1.4}_{-1.8} $&$-$\\
& Old &$17$&$>10^3$&$15 $ &PF &$-0.0^{+0.02}_{-0.02} $&$-$ & $-$ & $593^{+53}_{-83} $\\
\hline
\multirow{2}{*}{NGC 6503} &
Young &$22 $&$14 $ &$24 $ &PW &$-0.6^{+0.13}_{-0.16} $&$-0.2^{+0.12}_{-0.14} $&$28^{+1.5}_{-1.4} $&$-$\\
& Old &$18$&$21$&$17 $ &PF &$-0.4^{+0.10}_{-0.07} $&$-$ & $-$ & $284^{+203}_{-127} $\\
\hline
\multirow{2}{*}{NGC 7793} &
Young &$46 $&$20 $ &$72$&PW &$-1.5^{+0.23}_{-0.28} $&$-0.3^{+0.08}_{-0.08} $&$5.8^{+1.3}_{-1.3} $&$-$\\
& Old &$-$ & $-$ & $-$ & $-$ & $-$ & $-$ & $-$ \\
\hline
\bottomrule
\end{tabular}
\begin{tablenotes}
\small
\item \textbf{Notes}: Models compared: Single power law (S) - Equation~\ref{eq:ModelS}, PieceWise power law (PW) - Equation~\ref{eq:ModelPW}, and Single power law with exponential Fall-off (PF) - Equation~\ref{eq:ModelPF}. The parameters include: $\alpha_1$, the power-law slope common to all models, $\alpha_2$ the slope of the second power law in Model~PW, $\beta$ the transition point in arcsec ($\arcsec$) for Model PW, and $\theta_c$ the scale separation in arcsec of the exponential fall-off in Model~PF. The AIC values and the best-fit parameters are obtained from MCMC fits of the three aforementioned models to the observed TPCFs. Further description of the models, their parameters, and the fitting procedure are given in Section~\ref{sec:fitting}. The AIC is given by Equation~\ref{eq:AIC_eqn} and the model with the lowest AIC value is considered the best-fit model. This best-fit model is reported in column~6 ('Best Model') and only the best-fit parameters associated with that model are reported. Galaxies that had very few old clusters (i.e., NGC~3344, NGC~5253, and NGC~7793), and as a result very noisy TPCFs do not have AIC values or fits reported. In addition, in some cases a model has a very low likelihood, which leads to extremely high AIC values, in which case we denote the AIC value with a lower limit of $10^3$, i.e. $>10^3$. 
\end{tablenotes}
\end{threeparttable}
\end{table*}

\subsubsection{Model S: Single Power Law}
\label{sec:ModelS}

Many previous studies have reported pure power law TPCFs of the form described by model S, not just for star clusters, but also for individual stars and for tracers of gas \citep{Zhang_2001,Bastian_2005,Scheepmaker_2009,Gouliermis_2014,Gouliermis_2015,Gouliermis_2017,Grasha_2015,Grasha_2017_Spatial,Grasha_2018,Shabani_2018,Grasha_2019}.
We interpret power laws in the TPCFs as a sign of a self-similar hierarchical distribution in the star clusters. This is because the TPCF of a self-similar fractal distribution of points is a pure power law of the form $1+\omega(\theta) \propto \theta^{\alpha}$ \citep{Calzetti_1989,Larson_1995}, with the power-law slope $\alpha$ related to the two-dimensional fractal dimension $D_{2}$ as $D_2 = \alpha+2$. In Appendix~\ref{sec:appendix_fractals} we verify that toy fractal distributions show a pure power law TPCF up to the largest scale of the hierarchical structure, beyond which the TPCF flattens to approach a value of $1+\omega(\theta) \sim 1$. The amplitude $A_1$ of the power law (see Equation~\ref{eq:ModelS}) depends on both $D_2$ and the field of view length scale $R_s$, and agrees reasonably well with the analytical predictions of \citet{Calzetti_1988}. Physically, a self-similar hierarchy in the star clusters can originate from fractal density distributions in the natal star-forming gas, which in turn result from supersonic turbulent motions in the interstellar medium (ISM) \citep{Elmegreen_1993,Mac_2004,Kritsuk_2007,Federrath_Fractals}.  

\subsubsection{Model PW: Piecewise Power Law}
\label{sec:ModelPW}

The young clusters in some spiral galaxies show a piecewise power law TPCF. By examining the fits for these galaxies in Table~\ref{tab:AIC}, we can clearly see that the best-fit slope is steeper at smaller scales, and shallower at larger scales, i.e., $\alpha_1<\alpha_2$ (except in NGC~4449). The angular separation $\beta$ where the slope changes lies in the range \mbox{$\sim 2$--$14 \arcsec$}, which corresponds to scales of \mbox{$\sim 80$--$500 \, \mathrm{pc}$}. The break in slope cannot be edge effects, since it occurs at a scale well below the range at which we expect edge effects to play any role. Another possible cause for a break in the TPCF is a 2D-to-3D transition of the underlying distribution at the scale height of the galaxy. In other words, the distribution of star clusters could be three-dimensional at separations smaller than the scale height, but at separations beyond the scale height any pair of clusters would both lie on the plane of the galaxy, rendering the distribution two-dimensional. The projected fractal dimension, which is related to the slope of the TPCF, is different if we are looking at the projection of a thin slice (2D) rather than a thick disk (3D), and this projection effect leads to a change in the slope \citep[see,][for detailed models]{Sanchez_2008}. Observations of neutral hydrogen, Far-Infrared (FIR) dust, and $\gamma$-ray emission in nearby external galaxies are consistent with this mechanism \citep[][although see \citet{Koch_2020} for an alternative explanation for this transition]{Elmegreen_2001,Miville_2003,Ingalls_2004,Dutta_2009,Szotkowski_2019,Besserglik_2021}. However, our cluster TPCFs are not: as discussed in \citet{Sanchez_2010}, this transition should lead to a steeper slope at larger separations and a shallower ones at smaller separations, which is the opposite of what we find. 

Having ruled out edge effects and scale height effects, we conjecture that the breaks we see in galaxies with PW-type TPCFs represent real transitions from a fractal distribution at smaller scales set by turbulence to a mostly random distribution at larger scales where 2D galactic dynamics become more important than turbulence. Such a transition produces a shallow slope at large separation and a steeper slope at small separation, which is what we observe, and what is also seen in the correlation function of stars and \ion{H}{ii} regions in M33 \citep{Odekon_2008,Sanchez_2010}. In Section~\ref{sec:appendix_fractals}, we test this scenario by creating toy fractal distributions that are scale-free up to a maximum scale $L_{\mathrm{max}}$, and which are Poissonian at larger scales. We vary $L_{\mathrm{max}}$ in our models, attempt Model~PW fits to them, and find that the transition point parameter $\beta$ picks out the randomisation scale $L_{\mathrm{max}}$ quite well. Hence we infer Model~PW fits to represent fractal distributions that are scale-free only up to some maximum size scale $\sim \beta$, and become non-fractal at larger-scales. The match between the toy models and observations is not perfect, however: the measured TPCFs do not transition sharply to a completely flat slope like the toy models, but rather more smoothly to a value of $1+\omega \sim 1$. We speculate that this is because the distribution beyond the transition point $\beta$ is not \textit{entirely} Poissonian, since the large-scale distribution of clusters in a galaxy is clearly non-uniform on scales approaching the galactic scale length. Indeed, in the following section, we find direct evidence for this effect.

\subsubsection{Model PF: Power law with exponential cutoff}
\label{sec:ModelPF}

The third class of model, i.e., a power law that smoothly transitions to an exponential, is the best fit for the old clusters in all galaxies where we have enough old clusters to carry out a fit, and is also the best fit for young clusters in some of the dwarf galaxies in our sample. Following \citet{Mao_2015}, we hypothesise that this functional form reflects the large-scale distribution of the clusters in a galaxy, to which the young clusters converge as they age. To test this hypothesis, we distribute clusters in our toy models using a standard large-scale distribution: a radially thin exponential disk with a given scale length, and a Gaussian distribution in the vertical $z$ direction with a characteristic scale height. While there are more detailed models to describe the large-scale distribution of clusters in a galaxy, we chose the exponential disk model for simplicity. The toy model and its parameters are described in Appendix~\ref{sec:appendix_exponential}. We find that the TPCFs of the toy models display a smooth fall-off with separation, with a Model~PF fit yielding $\theta_c$ that corresponds to the exponential scale radius $r_c$ of the radial distribution (see, right panel of Figure~\ref{fig:exponential_TPCF}). Adding logarithmic spiral arms in the azimuthal direction to the exponential disk does not significantly change the TPCF. Since there is a close resemblance between the observed TPCFs and these toy models, we conjecture that PF-type TPCFs simply reflect the large-scale radial distribution in galaxies, which is reasonably well-described by a thin exponential disk with radial scale length $r_c$ approximately equal to the fitted scale length $\theta_c$. 


\subsection{Evolutionary Changes in the TPCF}
\label{sec:tpcf_agecomparison}
 
We have shown that the TPCFs for young ($<10$ Myr) and old ($>10$ Myr) clusters are significantly different. In most galaxies, older clusters have flatter, lower amplitude TPCFs, indicating that both they are less clustered overall, and that, unlike young clusters, older clusters do not follow a scale-free fractal structure. While a number of authors have reported qualitatively similar results, \citep{Odekon_2006,Sanchez_Alfaro_2009,Sanchez_Alfaro_2010,Grasha_2015,Grasha_2017_Spatial,Grasha_2018,Grasha_2019}, limited sample sizes have made it difficult to follow the evolution of the TPCF with time in detail. Because our sample of star clusters is among the largest available for this type of analysis, we can, at least for some galaxies, bin by age much more finely, and thereby obtain a higher resolution picture of TPCF evolution. We therefore divide star clusters into four age brackets: $T<2\, \mathrm{Myr}$, $2<T<10 \, \mathrm{Myr}$,
$10<T<100 \, \mathrm{Myr}$ and $T>100 \, \mathrm{Myr}$, and compute the TPCF for the distribution of clusters in these age brackets. We then use the fitting method outlined in Section~\ref{sec:fitting} to choose the best-fit functional form that describes the TPCF. We can only carry out this analysis for a subset of galaxies in our sample, as the rest do not have enough clusters at a wide enough range of ages.

\begin{figure}
    \centering
    \includegraphics[width=0.48\textwidth]{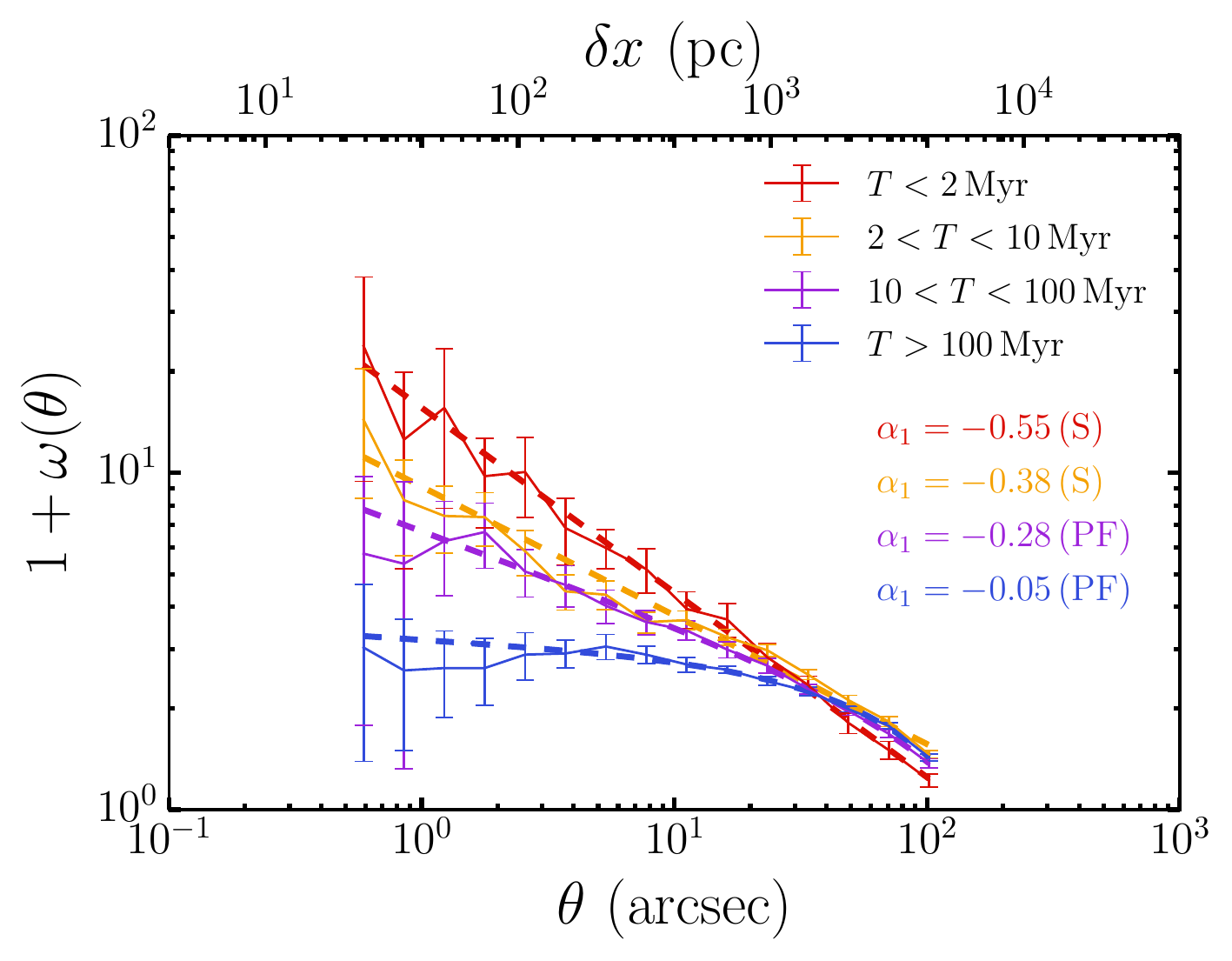}
    \caption{TPCFs (error bars with solid lines) with cluster age $T$ for NGC~5194, calculated in four age groups: $T\leq 2 \, \mathrm{Myr}$ (red), $2 < T \leq 10 \, \mathrm{Myr}$ (orange), $10 < T \leq 100 \, \mathrm{Myr}$ (violet), and $T>100 \, \mathrm{Myr}$ (blue). The best-fit model for each age group is overplotted with dashed lines, and the model name and parameters for each curve reported in the legend, both coloured by age group. It is evident that the TPCFs of younger clusters are scale-free power laws that decrease in slope as the clusters age. Older clusters are distributed more evenly across the disk and hence show shallower power-law slopes at small scales, with an exponential fall-off at large scales due to the correlation imposed by the overall radial scale in the disc.}
    \label{fig:AgeProgression_NGC5194}
\end{figure}

We show the result of this analysis in Figure~\ref{fig:AgeProgression_NGC5194} for NGC~5194, which has the highest number of clusters among all the galaxies in this study. We find similar qualitative behaviour for NGC~1313, NGC~1566 and NGC~0628, the other galaxies in our sample for which we were able to perform this analysis, albeit with substantially larger uncertainties due to the smaller numbers of clusters available. In the case of NGC~5194, we find that the best-fit model for clusters in the younger two categories are clearly single power laws (Model~S) with a slope $\alpha_1$ that decreases from $-0.55$ for the youngest clusters ($T<2\, \mathrm{Myr})$ to $-0.38$ for clusters with $2<T<10 \, \mathrm{Myr}$. We do not see any signs of an exponential fall-off from the exponential disk distribution for the younger clusters, or a break in the power-law that might indicate an outer limit to the scale-free structure. This is consistent with the visual impression from Figure~\ref{fig:cluster_positions}, which shows that younger clusters are mostly concentrated in hierarchically structured patterns that predominantly seem to trace the spiral arms in spiral galaxies, or the central regions of dwarf galaxies. On the other hand, clusters with ages $10<T<100 \, \mathrm{Myr}$ are best fit by a power law with an exponential fall-off at large scales (Model~PF), and a shallow power-law slope of $\alpha_1 = -0.28$ at small scales. This marks a transition phase where clusters are losing their natal fractal structure and thus have a shallower power law. However, these clusters are also old enough that they are distributed fairly uniformly across the extent of the disc, such that the imprint of the overall exponential radial distribution becomes evident at larger scales. Finally, the oldest clusters in the galaxy ($T>100 \, \mathrm{Myr}$) have a negligible small-scale slope $\alpha_1 \approx 0$, and a clear exponential fall-off at large scales, suggesting that for this age group fractal structure at all scales is completely lost, and the only remaining contribution to the TPCF comes from the large-scale radial structure of the disc.

This result is consistent with the finding in \citet{Grasha_2019} that star clusters in NGC~5194 become spatially decorrelated from molecular clouds by ages of \mbox{$\sim 50$--$100 \, \mathrm{Myr}$}. While our results on the TPCF in other galaxies are too noisy for us to perform a similar measurement in them, we note that \citet{Grasha_2018} found a lower cluster-molecular cloud decorrelation time in NGC-7793. Thus it is likely that the cluster-cluster decorrelation time that we are measuring will also depend on the host galaxy and its environment.

\subsection{Inferred Physical Properties of the Distribution and their Variation}
\label{sec:InferredProps}

\begin{figure}
    \centering
    \includegraphics[width=0.48\textwidth]{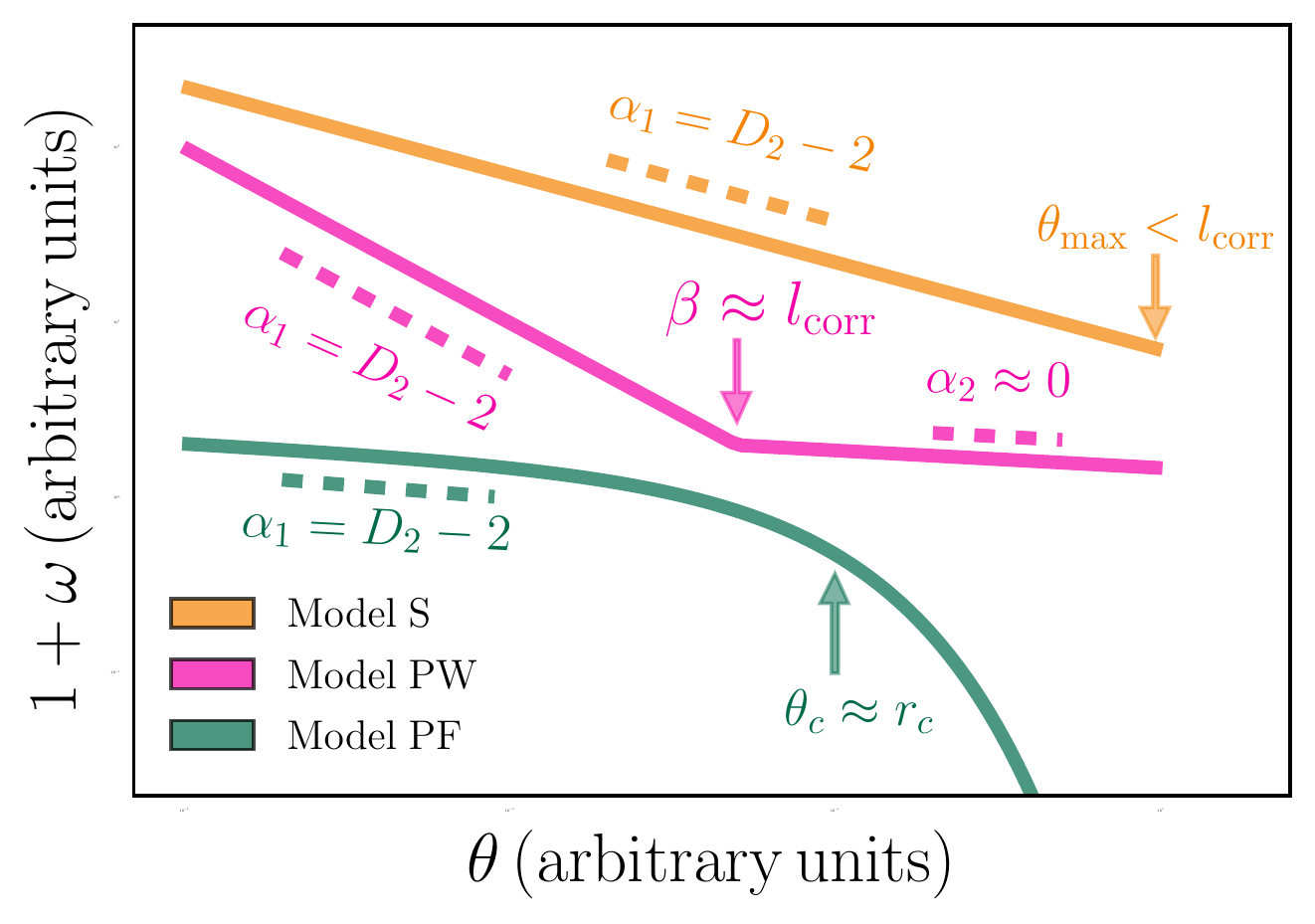}
    \caption{Schematic summarising the 3 functional forms fitted to the TPCF and the physical quantities that can be inferred from them, obtained based on our analysis using toy model distributions outlined in Section~\ref{sec:Features}. The 3 functional forms fitted are: Single power law (Model~S: Eq~\ref{eq:ModelS}), Piecewise power law (Model~PW: Eq~\ref{eq:ModelPW}), and Power law with exponential Fall-off (Model~PF: Eq~\ref{eq:ModelPF}). $\alpha_1$ is the small length-scale power-law slope in all three models, $\beta$ is the transition scale in Model~PW, $\theta_c$ is the exponential scale separation of Model~PF, and $\theta_{\mathrm{max}}$ is the largest bin to which the TPCF is measured. From these models and their best-fit parameters, we infer values of $l_{\mathrm{corr}}$, the largest scale to which hierarchical structure extends (see Section~\ref{sec:lcorr}), $D_2$, the 2D fractal dimension of the fractal distribution (see Section~\ref{sec:fractal_dimension}) - both calculated from the young cluster TPCF - and $r_c$, the exponential scale radii of the large-scale distribution in the galaxy, which is calculated from fits to the old clusters TPCF. Our estimate for $l_{\mathrm{corr}}$ is taken to be $\approx \beta$ from Model~PW, and $\ga \theta_{\mathrm{max}}$ from Model~S. Our estimate of $D_2 = \alpha_1+2$ for all three models, and that for $r_c$ is estimated to be $\approx \theta_c$. }
    \label{fig:toy_model_inferences}
\end{figure}

The three functional forms that we find provide a good description of the cluster TPCFs -- models S, PW, and PF -- and are characterised by three parameters: the largest scale up to which there is fractal signatures in the distribution $l_{\mathrm{corr}}$, the 2D fractal dimension of the distribution $D_2$ in the range of separations up to $l_{\mathrm{corr}}$ , and the scale radius $r_c$ beyond which the TPCF declines exponentially. We provide a schematic summary of these quantities, and their relationship to our functional forms, in Figure~\ref{fig:toy_model_inferences}. In the remainder of this section we investigate the distribution and variation of each of these quantities over the galaxy sample, and discuss possible physical origins for their values.

\begin{table}
\caption{Inferred physical quantities from TPCF fits.}
\centering
\label{tab:Inferred_Props}
\begin{threeparttable}
\begin{tabular}{l c c c}
\toprule
\multicolumn{1}{l}{Galaxy} & \multicolumn{1}{c}{$D_{\mathrm{2}}$} &  \multicolumn{1}{c}{$l_{\mathrm{corr}}$} & \multicolumn{1}{c}{$r_\mathrm{c}$}\\ 
\multicolumn{1}{l}{} & \multicolumn{1}{c}{} &  \multicolumn{1}{c}{[pc]} & \multicolumn{1}{c}{[kpc]}\\ 
\midrule
NGC 0628 &$0.9^{+0.14}_{-0.14}$ &$190^{+70}_{-40}$ &$18^{+3}_{-3}$\\
NGC 1313 &$1.4^{+0.03}_{-0.03}$ &$>960$ &$1.9^{+0.2}_{-0.2}$\\
NGC 1566 &$1.5^{+0.02}_{-0.02}$ &$>1730$ &$5.3^{+2.9}_{-1.6}$\\
NGC 3344 &$0.6^{+0.32}_{-0.37}$ &$110^{+100}_{-30}$ &$-$\\
NGC 3627 &$1.6^{+0.04}_{-0.04}$ &$>2020$ &$20^{+3.2}_{-4.0}$\\
NGC 3738 &$1.9^{+0.04}_{-0.09}$ &$-$ &$0.5^{+0.1}_{-0.1}$\\
NGC 4449 &$1.5^{+0.03}_{-0.03}$ &$<1440$ &$1.3^{+0.14}_{-0.13}$\\
NGC 5194 &$1.6^{+0.01}_{-0.01}$ &$>2700$ &$12.2^{+2.2}_{-1.6}$\\
NGC 5253 &$1.4^{+0.34}_{-0.37}$ &$<410$ &$0.6^{+1.1}_{-0.2}$\\
NGC 5457 &$1.4^{+0.11}_{-0.17}$ &$450^{+160}_{-200}$ &$19^{+1.9}_{-2.7}$\\
NGC 6503 &$1.4^{+0.13}_{-0.16}$ &$845^{+407}_{-244}$ &$8.7^{+6.2}_{-3.9}$\\
NGC 7793 &$0.5^{+0.23}_{-0.28}$ &$101^{+30}_{-25}$ &$-$\\
\bottomrule
\end{tabular}
\begin{tablenotes}
\small
\item \textbf{Notes}: See Figure~\ref{fig:toy_model_inferences} for a schematic outlining the method we use to obtain the values above from the fits in Table~\ref{tab:AIC}. $D_{\mathrm{2}}$: 2D fractal dimension inferred from the TPCF of young clusters, with error bars obtained from the fits.
$l_{\mathrm{corr}}$: Largest scale of hierarchical structure inferred from the young cluster TPCF. Error bars for $l_{\mathrm{corr}}$, if any, take into account the uncertainty in the distance to the galaxy and the fit uncertainty. The estimates of $l_{\mathrm{corr}}$ for galaxies where the TPCF is best fit by Model~S are lower limits. We do not report a value of $l_{\mathrm{corr}}$ for NGC~3738 since it shows no evidence of fractal structure ($D_2 \sim 2$). $r_\mathrm{c}$: Exponential scale radii derived from the TPCF of older clusters. Error bars take into account distance and fit uncertainties. We do not report $r_c$ values for NGC~3344 and NGC~7793, due to their lack of older clusters. For the same reason, for NGC~5253, $r_c$ is obtained from the Model~PF fit to the young cluster TPCF.
\end{tablenotes}
\end{threeparttable}
\end{table}

\subsubsection{Largest Scale of Hierarchical Structure}
\label{sec:lcorr}
The largest scale of hierarchical structure, $l_{\mathrm{corr}}$, denotes the maximum separation up to which star clusters are distributed in a scale-free fractal distribution, and beyond which star clusters are uncorrelated with each other. Since star clusters form from the underlying gas distribution in the ISM, we expect that $l_{\rm corr}$ is also approximately the size of the largest coherent gas structures \citep{Efremov_1995}. We infer $l_{\mathrm{corr}}$ from the TPCFs of the young clusters ($T \la 10 \, \mathrm{Myr}$), since these have been least influenced by evolutionary effects, and thus should most closely reflect the distribution at cluster formation. The method we use to estimate this scale for a galaxy depends on the best-fit model for its young cluster TPCF (see Table~\ref{tab:AIC}), and is summarised in the schematic shown in Figure~\ref{fig:toy_model_inferences}. For Model~PW galaxies, we take $l_{\rm corr}$ to be the TPCF transition point $\beta$ beyond which the distribution of clusters randomises. For Model~S galaxies, where the power law extends out to the last bin of measurement, we can only estimate a lower limit for this scale, taken as the largest scale for which we can measure the TPCF. For the dwarf galaxies, namely NGC~4449 and NGC~5253, we estimate $l_{\mathrm{corr}}$ as the scale where the power law sharply turns down due to the effect of the exponential disk distribution at larger scales (see Figure~\ref{fig:age_comparison}); this estimate is likely an upper estimate, as the effect of the exponential disk could be present even at smaller scales, and it is difficult to disentangle the power-law part from the exponentially falling part of the TPCF. We do not calculate a value of $l_{\mathrm{corr}}$ for NGC~3738 since it does not show any sign of scale-free fractal structure at any scale, since $\alpha_1$ is found to be $\sim 0$. The values of $l_{\mathrm{corr}}$ are reported in Table~\ref{tab:Inferred_Props}. 

\begin{figure*}
    \centering
    \includegraphics[width= 0.92 \textwidth]{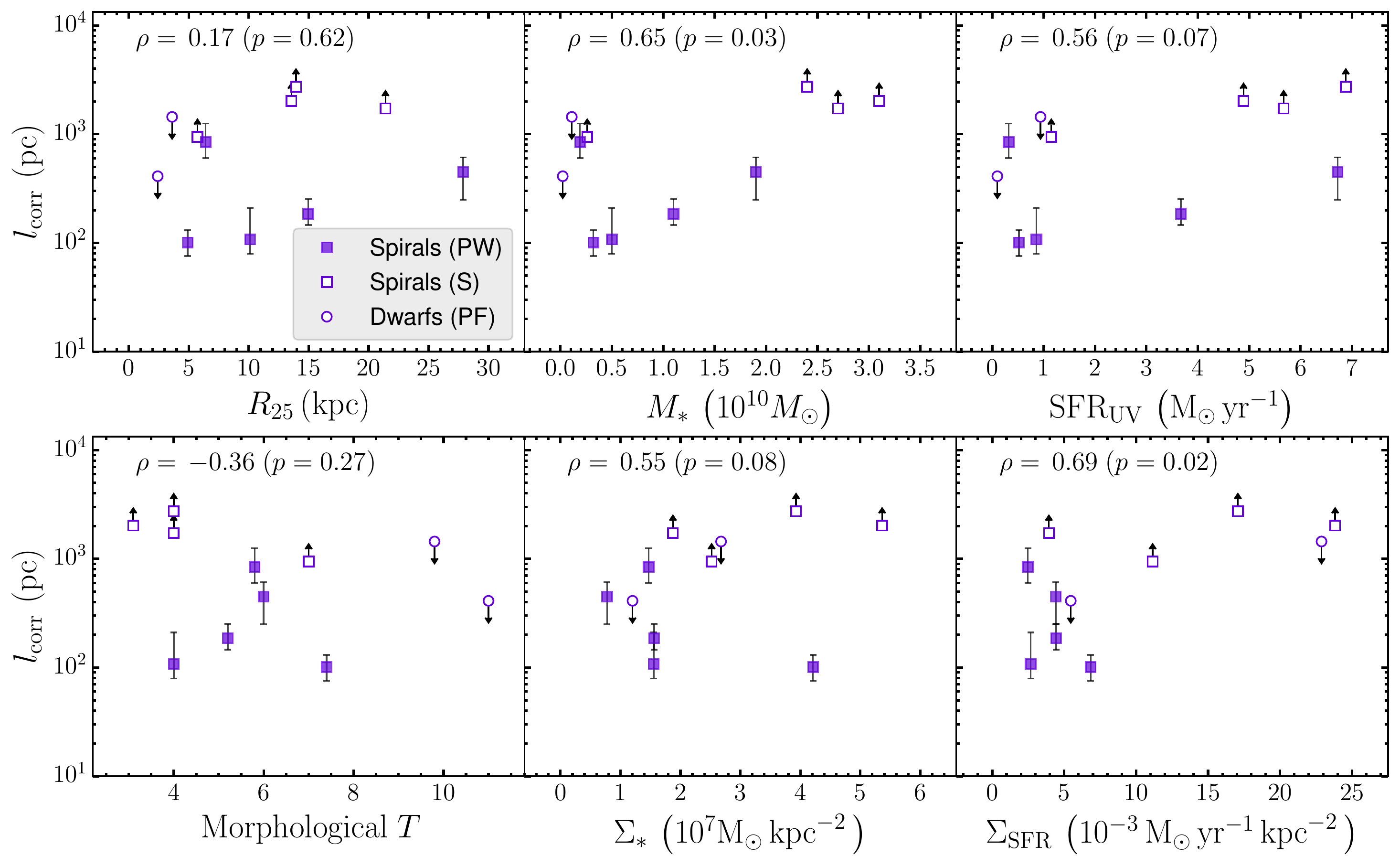}
    \caption{Comparison of the largest scale of hierarchical structure $l_{\mathrm{corr}}$ in star clusters reported in Table~\ref{tab:Inferred_Props} with the isophotal radius $R_{25}$, stellar mass $M_*$, UV-derived star formation rate $\mathrm{SFR}_{\mathrm{UV}}$, morphological $T$-value, stellar mass surface density $\Sigma_{*}$ and star formation rate surface density $\Sigma_{\mathrm{SFR}}$ of the host galaxy. The different marker styles denote the three ways that $l_{\mathrm{corr}}$ is estimated (see Section~\ref{sec:lcorr} and the schematic in Figure~\ref{fig:toy_model_inferences}) based on which functional form fits the young cluster TPCF best (reported in Table~\ref{tab:AIC}). The Pearson correlation coefficient $\rho$ and corresponding $p-$values of the correlation are provided for each pair of variables. We find signs of correlation of $l_{\mathrm{corr}}$ with $M_*$, $\mathrm{SFR}_{\mathrm{UV}}$, and $\Sigma_{\mathrm{SFR}}$, with stronger correlation if we restrict the sample to the spirals only. However, note the caveat that we use lower limits in the case of Model~S galaxies to calculate the values of $\rho$ and the associated $p-$values, and they may be different if we had constrained values of $l_{\mathrm{corr}}$ instead of lower limits.}
    \label{fig:lcorr_correlations}
\end{figure*}

As we can see, the values of $l_{\mathrm{corr}}$ vary among the galaxies and lie in a rather broad range from $\sim 100 \, \mathrm{pc}$ in NGC~7793 to upwards of $3000 \, \mathrm{pc}$ in NGC~5194. We compare these values with the various galaxy properties listed in Table~\ref{tab:galaxyinfo}: the standard isophotal radius $R_{25}$, the morphological $T$ type, the galaxy stellar mass $M_*$, the UV-derived star formation rate $\mathrm{SFR}_{\mathrm{UV}}$, and the stellar mass and star formation rate per unit area ($\Sigma_{*}$ and $\Sigma_{\mathrm{SFR}}$). We show scatter plots of $l_{\mathrm{corr}}$ against these quantities in Figure~\ref{fig:lcorr_correlations}, and report the Pearson correlation coefficients $\rho$ and their corresponding $p-$values in the Figure legend. The value of $\rho$ for a pair of variables lies in the range $-1$ to 1, with 1 ($-1$) indicating perfect linear correlation (anti-correlation) and 0 denoting no linear correlation; $p$ is the probability of obtaining a correlation coefficient $\geq \rho$ from a pair of variables that have, in fact, zero correlation (i.e., the null hypothesis), and is thus a measure of the statistical significance of the measured correlation. A value of $p<0.05$, meaning $<5\%$ probability of a false positive, is typically interpreted as statistically significant \citep[see, for instance,][]{Freedman2007}. We find moderately significant correlations of $l_{\mathrm{corr}}$ with $M_*$ ($\rho = 0.65$, $p=0.03$), $\mathrm{SFR}_{\mathrm{UV}}$ ($\rho = 0.56$, $p=0.07$), and $\Sigma_{\mathrm{SFR}}$ ($\rho = 0.69$, $p=0.02$), and no significant correlation with other quantities. The three detected correlations strengthen if one considers only the spirals in the sample. This analysis suggests that more massive and brighter galaxies (which also have higher star formation rates) tend to contain correlated complexes undergoing hierarchical star formation with larger sizes than are found in less massive galaxies, and agrees with similar signs of correlation found using the TPCF of star clusters in \citet{Grasha_2017_Spatial}. It is interesting to note that galaxy size ($R_{25})$ shows no correlation with $l_{\rm corr}$ ($\rho=0.17$, $p=0.62$), and that the area-averaged star formation rate ($\Sigma_{\mathrm{SFR}}$) correlates more strongly with $l_\mathrm{corr}$ than the total star formation rate $\mathrm{SFR}_{\mathrm{UV}}$. This suggests that $l_\mathrm{corr}$ is determined more by the physical conditions of the star-forming gas than by the overall size of the galaxy. However, we note that $\rho$ and the associated $p$-values are calculated using the lower (upper) limit value in the case of Model S (PF) fits, and hence might be different if we had real values. In addition, we caution that our sample is limited to only 12~galaxies, so any correlations are only suggestive, not conclusive, due to the low sample size.

\begin{figure}
    \centering
    \includegraphics[width= 0.44 \textwidth]{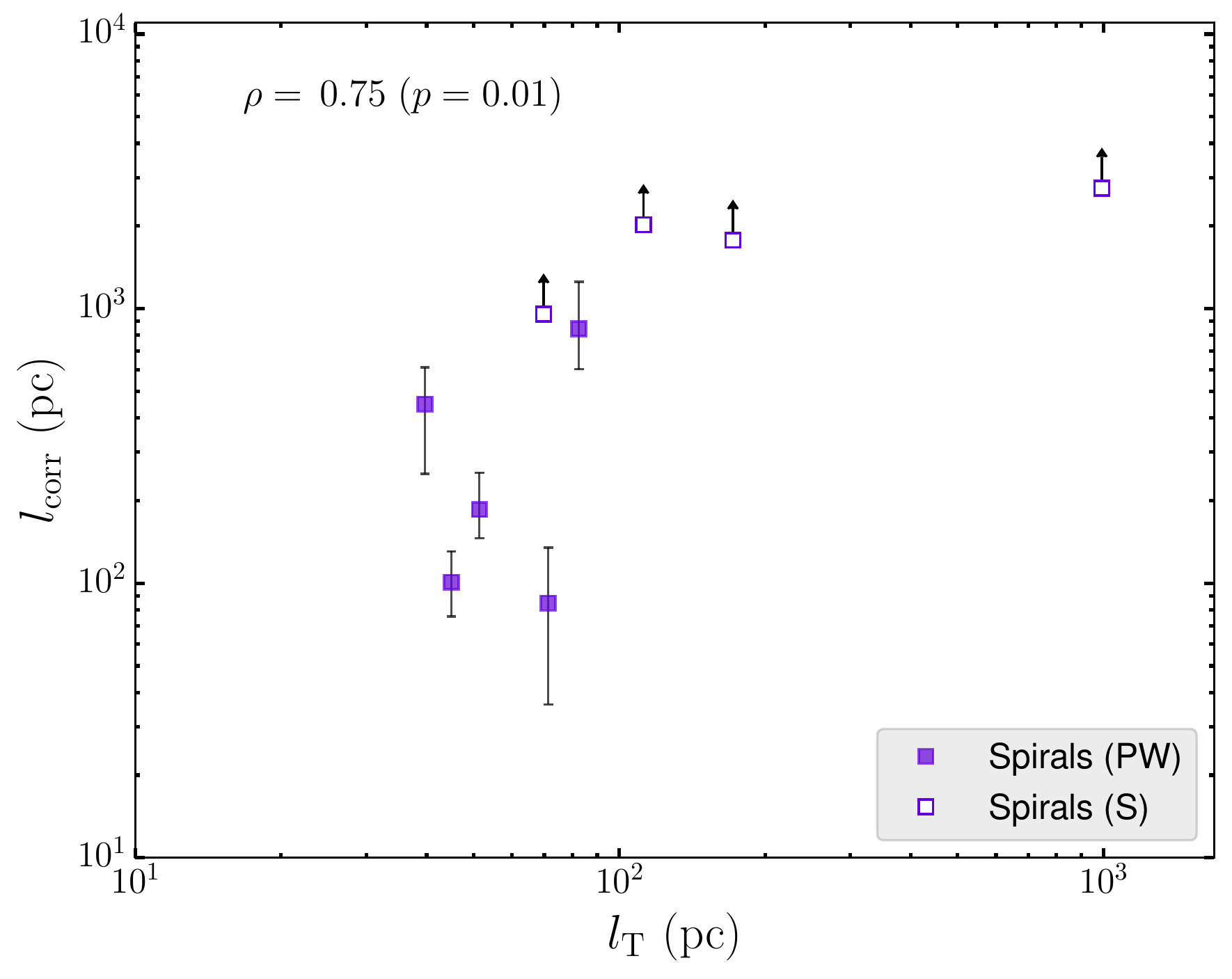}
    \caption{Comparison between the Toomre length $l_{\mathrm{T}}$ estimated using Equation~\ref{eq:toomre_length}, and the inferred largest scale of hierarchical structure in star clusters $l_{\mathrm{corr}}$ for the spiral galaxies in our sample. Marker styles are as outlined in Figure~\ref{fig:lcorr_correlations}. We find a statistically significant correlation with a Pearson correlation coefficient $\rho = 0.75$ and a $p-$value of $0.01$. This qualitatively suggests a physical picture where the largest scale of the hierarchy in star clusters is set by rotation-supported gravitational instability of the gas lying in the galactic disk.}
    \label{fig:Toomre_length}
\end{figure}

What physical mechanisms set $l_\mathrm{corr}$? The correlation with $M_*$ and $\Sigma_{\mathrm{SFR}}$ suggests that the gravitational potential of the matter (stars and gas) in the galaxy is important, with stronger potentials leading to larger complexes. This agrees with the physical picture of star-forming clouds at galactic scales being formed through gravitational instability in the disc, with their size ultimately limited by some \textit{top-down} mechanism that prevents them from growing too large. Galactic rotation is a candidate stabilising mechanism on large scales, and such a picture would be consistent with the results of \citet{Grasha_2017_Ages}, who find that a velocity gradient set by shear could explain the variation among the galaxies in the largest scale up to which pairs of star clusters are correlated in age. In this scenario, gravitational instability is unable to create structures past a certain maximum size, beyond which galactic rotation stabilises the disc. The natural scale in this case is the Toomre length $l_{\mathrm{T}}$ \citep{Toomre_1964,Escala_2008},
\begin{equation}
\label{eq:toomre_length}
    l_{\mathrm{T}} = \frac{4\pi^2G\Sigma_\mathrm{g}}{\kappa^2},
\end{equation}
where $G$ is the gravitational constant, $\Sigma_\mathrm{g}$ is the gas surface density, and $\kappa$ the epicyclic frequency of rotation. For a flat rotation curve, which we assume here, $\kappa = \sqrt{2}\Omega$, and $\Omega = v_{\mathrm{rot}}/r$ is the angular rotational velocity calculated from the flat rotational velocity $v_{\mathrm{rot}}$ at a given galactocentric radius $r$. To check this hypothesis, we compute $l_\mathrm{T}$ for all the spiral galaxies in our sample; we omit dwarf/irregular galaxies due to the lack of robust observed rotational curves as it is unclear to what extent the dwarfs have a disc-like structure.  This calculation requires estimates for $\Sigma_\mathrm{g}$ and $v_{\mathrm{rot}}$, and an appropriate choice for $r$. To calculate $\Sigma_\mathrm{g}$, we use galaxy-averaged molecular gas ($\mathrm{H}_2$) surface densities reported in the literature where available, and estimated total molecular gas masses $M_{\mathrm{H}_2}$ from the literature divided by $\pi R_{25}^2$ otherwise. We use \ion{H}{i} rotation curves and their reported rotational velocities available in the literature to infer $v_{\mathrm{rot}}$. We choose the representative radius $r$ at which to calculate $l_{\mathrm{T}}$ to be the median galactocentric radius of the young star clusters in our star cluster catalogues. We show $l_{\mathrm{T}}$ versus $l_{\mathrm{corr}}$ in Figure~\ref{fig:Toomre_length}. We find a reasonably strong ($\rho = 0.75$) and statistically significant ($p=0.01$) correlation between the two. The two scales we calculate, although correlated, are not identical; in general $l_{\mathrm{corr}}$ is larger than $l_{\mathrm{T}}$ by a factor of a few. We caution that modern treatments of the Toomre instability include the effects of multiple stellar populations along with the gas, the effects of finite thickness, and the dissipative nature of gas \citep[see, e.g., ][]{Romeo_2013}. However, we lack measurements of the stellar velocity dispersion or disc scale height, which would be required to include these effects, and thus we limit our comparison to the simple pure-gas Toomre length. It is important to extend this comparison to a larger sample of galaxies, to obtain more robust and conclusive results for the importance of such a mechanism. 

\subsubsection{Fractal Dimension of Young Clusters}
\label{sec:fractal_dimension}
Another quantity of interest is the fractal dimension of the hierarchical distribution at scales $l<l_{\mathrm{corr}}$. The fractal dimension is a quantity that characterises self-similar structure in a distribution, with lower values corresponding to less space-filling hierarchical structures. Self-similar hierarchies are proposed to be set self-consistently by interstellar turbulence in the ISM gas \citep{Elmegreen_2004,Federrath_Fractals}. If this picture is correct, the result should be a nearly universal value for the fractal dimension, as has been proposed in earlier studies \citep{Feitzinger_1987,Elmegreen_1996_Mass}. Previous studies of the fractal dimension of the gas and/or dust distribution in galaxies have generally been consistent with the hypothesis of a universal fractal dimension \citep[see, Table 1,][]{Shadmehri_2011}. However, \citet{Sanchez_2008} find statistically significant variation in the fractal dimension of $\ion{H}{ii}$ regions with host galaxy and/or environment. Here, we investigate whether the fractal dimensions of the distributions of star clusters in our sample are the same in all galaxies, and if not, how its variation correlates with other galactic properties.   

We compute the 2D fractal dimension $D_2$ from the power-law slope of the fits to the young cluster TPCFs reported in Table~\ref{tab:AIC}, using the relation $D_2 = 2+\alpha_1$, where $\alpha_1$ is the fitted slope of the power law. Since all 3 fit models include $\alpha_1$ as a parameter, we obtain a corresponding $D_2$ for all galaxies in our sample. This approach is summarised in the schematic shown in Figure~\ref{fig:toy_model_inferences}. As in the previous section, we do this for the young clusters TPCF, which should more closely reflect the fractal dimension of the natal gas supposedly set by interstellar turbulence. We list $D_2$ for each galaxy in our sample in Table~\ref{tab:Inferred_Props}. We find variations well beyond the computed $1-\sigma$ errors, with $D_2$ lying in the general range \mbox{$0.5$--$1.6$}, with the exception of NGC~3738, which has a value $D_2$ corresponding to a completely random distribution, i.e., $D_2 \sim 2.0$. This suggests that, consistent with \citet{Sanchez_2008}, and contrary to earlier suggestions \citep{Feitzinger_1987}, the hierarchical structuring in the star cluster distribution does not show signs of universality and depends on the host galaxy and its properties in a way that the gas distribution apparently does not \citep{Elmegreen_1996_Mass, Shadmehri_2011}. 

\begin{figure*}
    \centering
    \includegraphics[width=0.92 \textwidth]{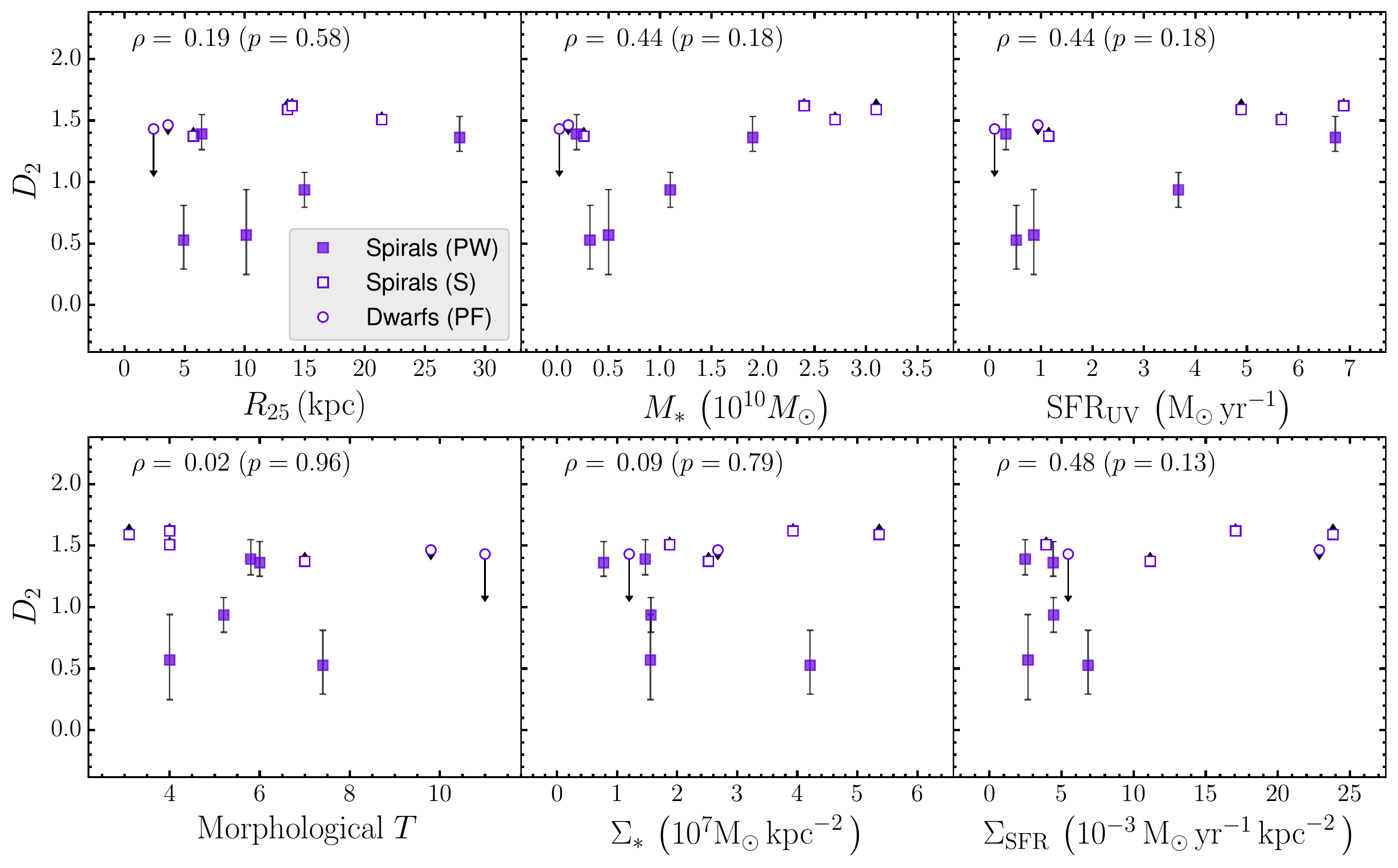}
    \caption{Same as Figure~\ref{fig:lcorr_correlations}, but for the inferred 2D fractal dimension $D_2$ of the young cluster distribution, except for NGC~3738 which has a value of $D_2$ corresponding to a Poissonian distribution. Note that some points have very small errors, which are not visible. Overall, we find weak signs of correlations for $D_2$, which are, however, not statistically significant.}
    \label{fig:D2_correlations}. 
\end{figure*}

We show scatter plots of $D_2$ versus various galaxy properties in Figure~\ref{fig:D2_correlations}; we report the Pearson correlation coefficient of for each of the comparisons shown in the corresponding Figure panels. As with $l_{\mathrm{corr}}$, we find at most marginal evidence for correlation of $D_2$ with $M_*$, $SFR_{\mathrm{UV}}$, and $\Sigma_{\mathrm{SFR}}$; the correlation is stronger if we consider only the spirals in the sample, but remains below the level of statistical significance. To the extent that we interpret the vague hints in our data, they suggest that more massive galaxies have larger fractal dimensions (more space-filling distributions) than less massive galaxies. Such a trend for the inferred fractal dimension have been reported in earlier studies - i.e., brighter galaxies - quantified by their $B-$band absolute magnitude - have higher fractal dimensions than fainter ones \citep{Parodi_2003,Odekon_2006,Sanchez_2008}. In addition, \citet{Sanchez_2008} found that this correlation disappears when the irregular galaxies are included in their analyses, as irregular galaxies have fractal dimensions similar to the brightest spiral galaxies, but are also significantly fainter then them, qualitatively similar to what we find. It would be interesting to search for a similar effect for clusters using a larger sample of galaxies. 

We also point out that there are earlier estimates for $D_2$ in the literature for a few of our galaxies. The values we obtain are consistent within the uncertainty in some galaxies, but not for all. For instance, \citet{Scheepmaker_2009} estimate $D_2 \sim 1.6$ for clusters younger than $\sim 30 \, \mathrm{Myr}$ in NGC 5194, which is consistent with our result ($1.6 \pm 0.1$). On the other hand, the values we obtain for NGC 0628 ($\sim 0.9$) and NGC 6503 ($\sim 1.4$) are different than earlier values quoted for them in literature - i.e., 1.5 for NGC 0628 \citep{Elmegreen_2006,Gusev_2014} and 1.7 for NGC 6503 \citep{Gouliermis_2015}. This difference could occur for several reasons. For instance, these studies do not look at the hierarchical structuring of star clusters, but rather star forming regions (in NGC 0628) or young stars (in NGC 6503), and there is no reason to assume that these structures all have the same fractal dimension. In addition, the NGC 0628 studies inferred a value of $D_2$ from the slope of the cumulative size distribution of star-forming regions, whereas we infer $D_2$ from the slope of the TPCF of young star clusters, a very different method. It is also well known that differential clustering estimates -- such as the TPCF -- are well suited to determining scales at which a change in clustering strength takes place \citep[see, for instance, ][]{Sanchez_Alfaro_2010}. This feature, combined with our Bayesian approach to fitting various functional forms and hence slopes is important, especially in the cases of NGC 0628 and NGC 6503, which were best-fit by Model~PW, and for which a fit to Model~S only (analogous to the procedures used in earlier work, which implicitly assume a single power law correlation function) would yield a significantly shallower slope, and hence a higher $D_2$. 

\subsubsection{Exponential Scale Radii}
\label{sec:Scale_Radii}
\begin{figure}
    \centering
    \includegraphics[width=0.48 \textwidth]{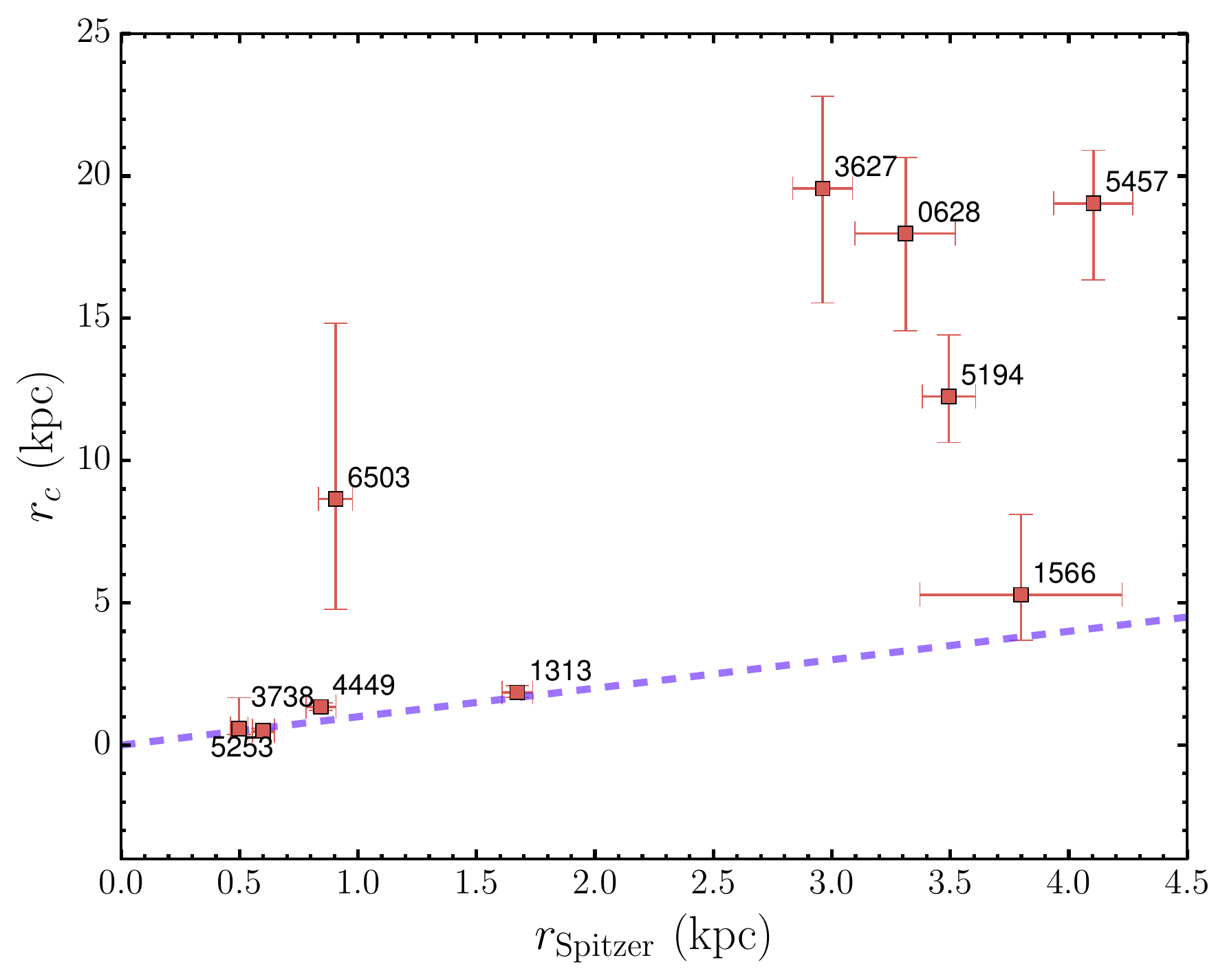}
    \caption{Comparison between the exponential scale radius inferred from the TPCFs $r_c$ (see Section~\ref{sec:Scale_Radii}) and $r_{\mathrm{Spitzer}}$, the value reported in the Spitzer Survey of Stellar Structure in Galaxies \citep[S$^4$G,][]{Salo_2015}. Error bars are plotted for our inferred value $r_c$, taking into account errors from the fit to the TPCF and the uncertainty in the distance to the galaxy. The error bars for $r_{\mathrm{Spitzer}}$ only take into account the uncertainty in the distance as \citet{Salo_2015} do not report error bars for their calculated scale lengths. A one-to-one dashed line (purple) is added to guide the eye. As we can see, $r_c$ reasonably reproduces $r_{\mathrm{Spitzer}}$ for smaller galaxies, but overestimates it for larger galaxies where the \textit{HST} field of view does not adequately cover the outer galaxy (see main text).}
    \label{fig:rc_comparison}
\end{figure}

The exponential scale angle $\theta_c$, corresponding to a linear distance $r_c$, is set by the radial distribution of clusters in the galaxy, and obtained by fitting Model~PW to the TPCF of old clusters ($T>10\, \mathrm{Myr}$), as indicated in the schematic shown in Figure~\ref{fig:toy_model_inferences}. We report values of $r_c$ in Table~\ref{tab:Inferred_Props}; note that the reported uncertainties include the uncertainty in the distance to the galaxy. We compare our $r_c$ values with $r_{\mathrm{Spitzer}}$ - the scale radius of the galaxies in our sample estimated with the $3.6$ and $4.5 \micron$ Spitzer Survey of Stellar Structure in Galaxies \citep[S$^4$G,][]{Salo_2015} in Figure~\ref{fig:rc_comparison}. We find reasonable agreement for galaxies that have lower values of $r_c$, especially the dwarfs, but for most larger galaxies we find $r_c \gg r_{\mathrm{Spitzer}}$. Why might this be the case? One possibility is that there are substantial uncertainties in $r_{\rm Spitzer}$, since S$^4$G provides no estimate of uncertainties apart from those arising from the distance uncertainty. However, this seems unlikely to account for the factor of $3-4$ discrepancy we find for large $r_c$. A more likely explanation is that the HST field-of-view does not encompass the entire extent of the disc as it does for the smaller galaxies. To test whether this could lead to overestimates of $r_c$, we artificially place a limited field-of-view on our toy galaxy models (see Appendix~\ref{sec:appendix_exponential}). We then compute and fit model~PF to the TPCFs, and check whether the value of $r_c$ derived from the fitted $\theta_c$ overestimates the true input value of the scale length we provide. In Figure~\ref{fig:Rmax_fits} we show that this is indeed the case: limiting the field of view to 2 galactic scale lengths leads to an overestimate of $r_c$ by a factor $\sim 3$, roughly the observed discrepancy. We therefore tentatively conclude that the exponential cutoff found in Model~PF gives a reasonable estimate of the scale length of the host galaxy, but only as long as the footprint within which the clusters are sampled extends to sizes significantly larger than the galactic scale length. 

\section{Summary}
\label{sec:Summary}
In this study, we investigate the hierarchical spatial distribution of young star clusters in 12~local galaxies catalogued with the LEGUS survey \citep{Calzetti_2015}, using the angular two-point correlation function (TPCF) $1+\omega(\theta)$ as a function of angular separation $\theta$. Our sample consists of various morphological types, from irregular dwarfs to grand design spirals, allowing us to probe the effects of the host galaxy environment on the star cluster distribution. Estimated ages for the clusters obtained as part of the survey also allow us to study how the cluster distribution changes with age. We show that the TPCFs in all our galaxies are reasonably well-described by a model characterised by three parameters: the largest scale of hierarchical structure $l_{\mathrm{corr}}$, the 2D fractal dimension of the young star cluster distribution $D_{\mathrm{2}}$, and the radial exponential scale radii $r_c$ of the star clusters. We study how these parameters vary with the properties of the galaxies, to investigate the physical mechanisms that might be responsible in setting them. Our main results are summarised below.

\begin{enumerate}
    \item The TPCFs of younger clusters show large correlation amplitudes and strong fractal structure characterised by scale-free power-law TPCFs for separations $\theta \la l_{\mathrm{corr}}$. The TPCFs of older clusters, on the other hand, show shallow power laws, characteristic of more randomised distributions at smaller separations, and an exponential fall-off at larger separations (Figure~\ref{fig:age_comparison}). Comparison with toy models shows that this fall-off is consistent with the cluster distribution following an overall exponential decline with galactocentric radius (Section~\ref{sec:ModelPF}).
    \item The star cluster distribution loses its natal hierarchical structure gradually with age (Figure~\ref{fig:AgeProgression_NGC5194}), with the TPCF successively flattening as the age of the population increases, occupying a larger extent of the disc, and eventually converging to the residual correlation from the large exponential disc distribution in the galaxy. 
    \item We find a range of values of $l_{\mathrm{corr}}$ across the sample, from $\sim 100 \, \mathrm{pc}$ to scales beyond $\sim 2.5 \, \mathrm{kpc}$, the largest we can reliably measure given the size of the LEGUS footprint. Similarly, we find a range of fractal dimensions ($D_2$) for young clusters from \mbox{$\sim 0.5$--$1.9$} across our sample of galaxies (see Table~\ref{tab:Inferred_Props}). The range of these parameters is substantially larger than the uncertainties, and suggests that there are significant variations in the hierarchical structuring of star clusters from one galaxy to another. Earlier studies show that this is not the case for the gas distribution \citep[see, Table 1,][]{Shadmehri_2011}, suggesting that there might be additional physical mechanisms at play in explaining these differences. 
    \item We find signs of some positive correlation of $l_{\mathrm{corr}}$ with stellar mass $M_*$, UV-derived star formation rate $\mathrm{SFR}_{\mathrm{UV}}$ and star formation rate surface density $\Sigma_{\mathrm{SFR}}$ (Figure~\ref{fig:lcorr_correlations}). We also find relatively stronger and statistically more significant correlation of $l_{\mathrm{corr}}$ with the galaxy-averaged Toomre length $l_{\mathrm{T}}$ in the disc (Figure~\ref{fig:Toomre_length}), suggesting that rotation-supported gravitational instability might be an important mechanism in setting the scales where gas is hierarchically structured. We stress, however, that we are limited to 12~galaxies in this study, and hence, cannot make fully conclusive inferences.
    \item We demonstrate that we can robustly infer an estimate for the radial scale length of the star cluster distribution in the galaxy ($r_c$) from the TPCF of its more randomly distributed older clusters, but only for galaxies where the field of view within which we measure star cluster positions is substantially larger than the radial scale length (Figure~\ref{fig:rc_comparison}). 
\end{enumerate}

Overall, our results suggest that the hierarchical structure of star clusters, both old and young, is not universal, but instead depends on the physical properties of the host galaxy. For older clusters this dependence is relatively trivial, since as the cluster population ages, it loses the hierarchical structure with which it formed, and the resulting TPCF simply reflects the overall size of the galaxy. More intriguingly, though, even for young clusters we measure statistically significant variations in both the fractal dimension and the largest scale of the hierarchical distribution, and show that these correlate with large-scale galactic properties. Therefore, cluster formation is possibly not a universal process that operates the same way in all galaxies, which suggests significant scope for future work by extending our study to a larger sample, within which the correlations between cluster distributions and galactic properties of which we see hints can be more reliably measured.

\section*{Acknowledgements}
We thank the anonymous referee for a constructive review that improved the quality of the paper. S.~H.~M.~would like to thank Dimitrios Gouliermis, James Beattie and Piyush Sharda for insightful discussions during the course of the project. Based on observations made with the NASA/ESA Hubble Space Telescope, obtained at the Space Telescope Science Institute, which is operated by the Association of Universities for Research in Astronomy, Inc., under NASA contract NAS 5-26555. These observations are associated with program \# 13364. C.~F.~acknowledges funding provided by the Australian Research Council through Future Fellowship FT180100495, and the Australia-Germany Joint Research Cooperation Scheme (UA-DAAD). M.~R.~K.~acknowledges funding from the Australian Research Council through its \textit{Discovery Projects} and \textit{Future Fellowship} funding schemes, awards DP190101258 and FT180100375. M.~M.~acknowledges the support of the Swedish Research Council, Vetenskapsrådet (internationell postdok 2019-00502). This project has received funding from the European Research Council (ERC) under the European Union's Horizon 2020 research and innovation programme (grant agreement No 757535). The following software tools were used for analysis and plotting : \verb|NumPy| \citep{numpy}, \verb|SciPy| \citep{scipy}, \verb|Matplotlib| \citep{matplotlib}, and \verb|corner| \citep{corner}. Some of the figures used in this work make use of the \verb|CMasher| package \citep{Cmasher}. This research made use of Astropy,\footnote{\url{http://www.astropy.org}} a community-developed core Python package for Astronomy \citep{astropy_2013,astropy_2018}. This research has also made use of NASA's Astrophysics Data System (ADS) Bibliographic Services. 


\section*{Data Availability}
The data (cluster catalogues) underlying this article are publicly available online on the Mikulski Archive for Space Telescopes \footnote{\url{https://archive.stsci.edu/prepds/legus/dataproducts-public.html}} (MAST)  for all galaxies except NGC 3627 and NGC 5457, which will be published in a forthcoming paper (Linden et al., in prep). The tools and code to reproduce the analysis and plots of the paper are publicly available at \url{https://github.com/shm-1996/legus-tpcf}.



\bibliographystyle{mnras}
\bibliography{TPCF_Legus} 




\appendix




\section{Edge Effect Quantification}
\label{sec:appendix_edge}

\begin{figure}
    \centering
    \includegraphics[width=0.48\textwidth]{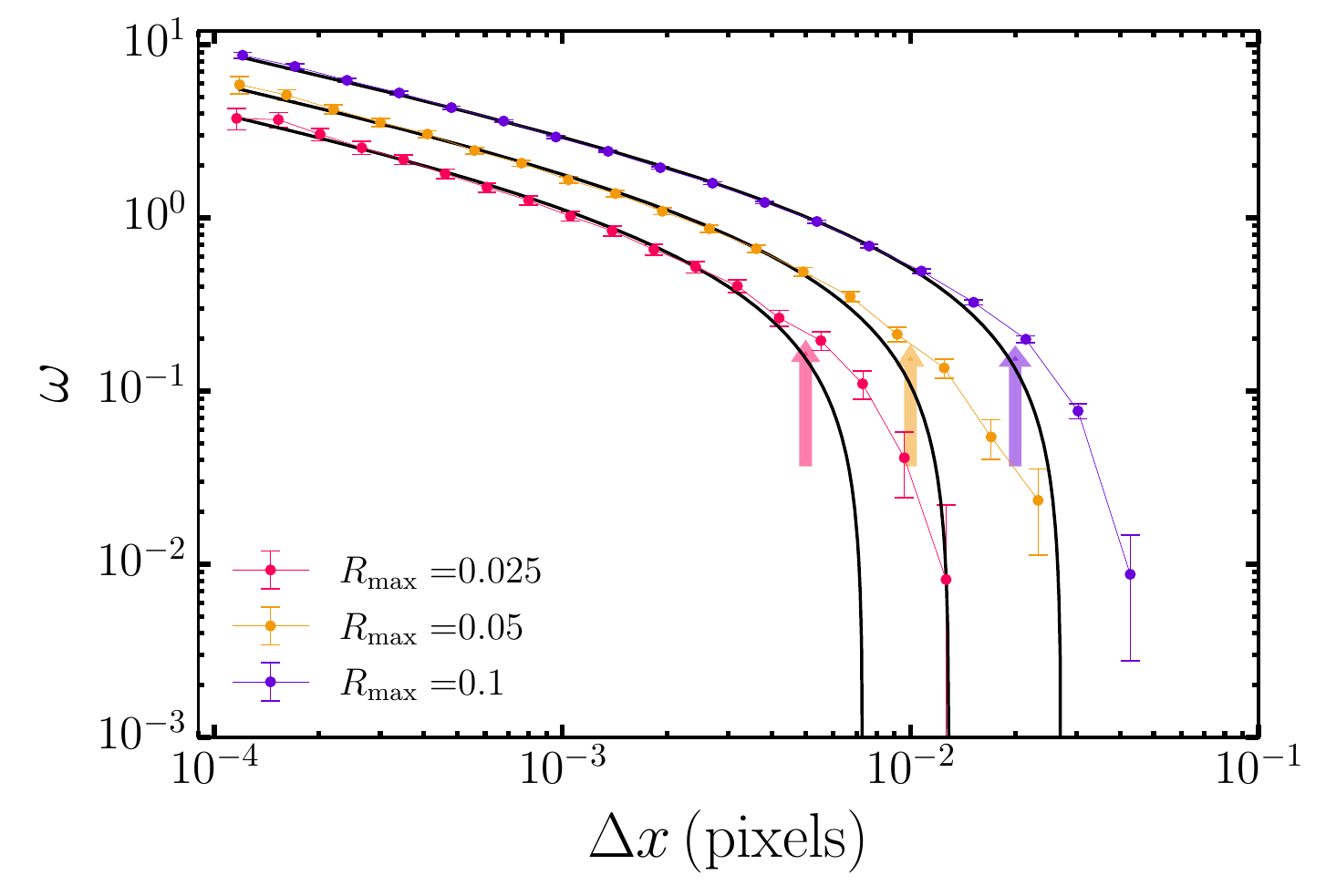}
    \caption{TPCFs as a function of separation $\Delta x$ for limited field-of-view squares of size $R_{\mathrm{max}} = 0.025,0.05$ and $0.1$, placed randomly on a fractal distribution with $D_2=1.5$ whose scale-free behaviour extends up to $L \sim 0.125$. Solid lines in black show the analytical relation for the TPCF of a fractal $\omega = A(\Delta x)^{-0.5}-1$, as expected from \citet{Calzetti_1988}. We find that the calculated and analytical TPCFs match for scales up to $\Delta x \sim R_\mathrm{max}/5$, beyond which there is significant deviation. We denote this scale by the arrows shown in the plot.}
    \label{fig:EdgeEffect}
\end{figure}

In this section we quantify the minimum scale beyond which edge effects caused by a limited field-of-view might play a role in determining the value of the TPCF, as discussed briefly in Section~\ref{sec:edgeeffects}. We do this by preparing pure fractal distributions with known TPCFs (see Section~\ref{sec:appendix_fractals}, and \citealt{Calzetti_1988}) that show scale-free TPCFs up to a length scale $L \sim 0.125$. We then truncate the distribution to a square field-of-view with side lengths $R_{\mathrm{max}}<L$, which causes deviations from the input TPCFs at separations $\Delta x >r_{\mathrm{edge}}$ where $r_{\mathrm{edge}}$ is the minimum length scale where edge effects start to play a role. By measuring where the TPCFs of our truncated distributions deviate from the input one, we obtain a measurement of $r_{\rm edge}$. We carry out this experiment for $R_{\mathrm{max}} = 0.025,0.05$ and $0.1$. For each value of $R_{\mathrm{max}}$, we compute the mean TPCF over 30 different realisations obtained by placing the field-of-view square at random locations in the fractal. For our input pure fractals, the expected TPCF is $1+ \omega(\theta) = A\theta^{\alpha} $, where $\alpha = D_2+2$ and $D_2$ is the 2D fractal dimension of the distribution, and the normalisation $A$ depends on a field of view length scale or radius $R_{\mathrm{max}}$ as $A = (1+\alpha/3)R_{\mathrm{max}}^{-\alpha}$. We fit this analytical form to our measured TPCFs and investigate at what point our fitted values of $\alpha$ and $A$ differ from the values for the input, non-truncated fractal distribution by more than 10\%. We show our computed TPCFs from the truncated data along with the true TPCFs for each value of $R_{\mathrm{max}}$ in Figure~\ref{fig:EdgeEffect}. We plot $\omega$ instead of $1+\omega$ in order to make the edge effects more clearly visible.

In general, we find that our TPCFs for the truncated data match analytic expectations to better than 10\% for separations $\Delta x \sol R_\mathrm{max}/5$, but that for the truncated-data TPCFs $\omega$ falls off much more shallowly than predicted by the analytical relation. This disagreement is likely due to the data-random cross correlation term of the \citet{Landy_1993} estimator, which becomes dominant at separation close to the size of the field-of-view. Given this result, we set $l_\mathrm{edge} = R_{\mathrm{max}}/5$, and discard our measured TPCFs at larger separations. However, we caution that our choice $R_\mathrm{max}/5$ is somewhat arbitrary, since the divergence between the measured and true TPCFs in our idealised experiment occurs over a finite range of scales, rather than sharply at a single scale.

\section{TPCF of All Clusters}
\label{app:TPCF_all}
\begin{figure*}
    \centering
    \includegraphics[width=\textwidth]{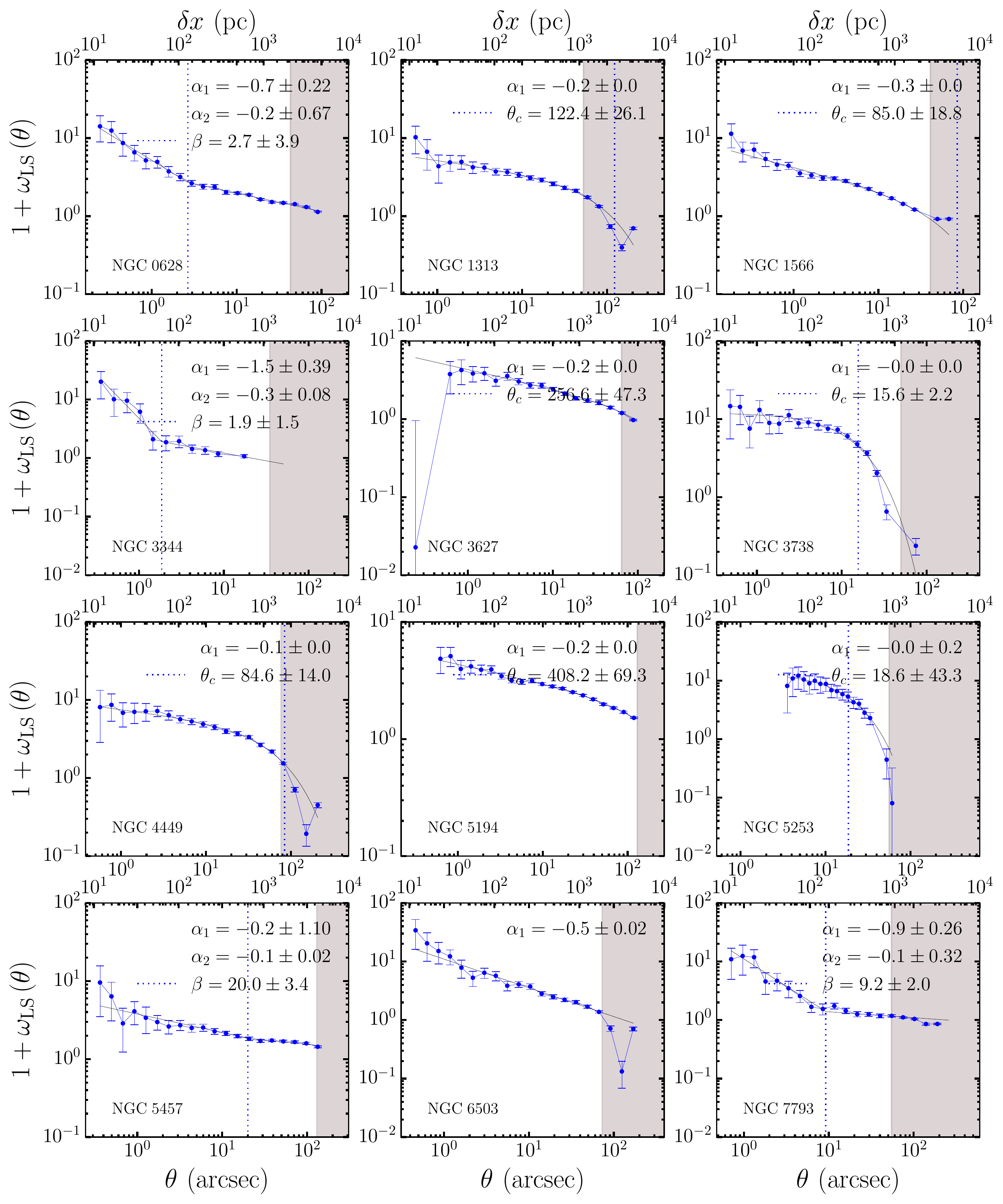}
    \caption{TPCF of star clusters of all ages for each galaxy in the sample, with their best-fit models overplotted, using the approach outlined in Section~\ref{sec:fitting}. The best-fit parameters appropriate to the best-fit models are denoted on the plot, and the grey shaded regions denote separations where edge effects might play a role, as in Figure~\ref{fig:age_comparison}.}
    \label{fig:tpcf_all}
\end{figure*}

In Section~\ref{sec:tpcf_galaxies} we presented and discussed the TPCF of star clusters divided into young and old clusters based on an age cut ($T = 10 \, \mathrm{Myr}$). We chose this approach instead of showing the combined TPCF of both young and old clusters, as the physically relevant features in the TPCF are more clearly evident when the sample is divided by age. For completeness, however, we show the combined TPCF in Figure~\ref{fig:tpcf_all} with their best-fit models obtained from the methodology outlined in Section~\ref{sec:fitting} overplotted. 

\section{Toy Models}
\label{sec:appendix_toymodels}
In this section we describe a set of physically motivated toy models that we use to infer the features seen in the star cluster TPCFs of the galaxies in our sample in Section~\ref{sec:tpcf_galaxies}. These three models are meant to characterise the three fitting functional forms described in Section~\ref{sec:fitting}, namely a single power law (Model~S), a piecewise power law (Model~PW) and a power law with an exponential fall-off (Model~PF). We explain the three classes of features by a pure fractal distribution, fractal distribution that transitions to a random one beyond some outer scale, and a radially exponential disc distribution, respectively. Below we discuss the parameters of the model, and how the TPCFs of the model depend on the parameters.

\subsection{Fractal Distributions}
\label{sec:appendix_fractals}

Our procedure for constructing fractal distributions of points uses the same reverse box-counting method previously employed by a number of authors \citep{Bate_1998,Cartwright_2004,Gouliermis_2014,Elmegreen_2018}. The method is as follows: we begin with a square of side length $L_{\rm box}$, which we divide up into $2^l$ square cells of side length $L_\mathrm{box}/2^l$; where $l\geq 0$ is the level in the hierarchy. We start at a base level $l_{\rm base}$ by marking all cells at that level as ``active''. We then subdivide each active cell into four sub-cells at level $l = l_{\rm base}+1$, and randomly decide whether to mark those sub-cells as active, with probability $p = 2^{D_{\mathrm{2}} - 1}$. We then repeat this procedure recursively: for each active cell at level $l$, we subdivide it into four cells and level $l+1$, which we mark active or inactive with probability $p$, and so forth. The algorithm terminates at some predetermined maximum level $l_{\rm max}$; we place a point in each active cell on this level, with the location of the point set equal to the location of the cell centre plus a small random dither to avoid an overly-gridded structure.

\begin{figure*}
    \label{fig:Fractal_D}
    \centering
    \includegraphics[width=0.98\textwidth]{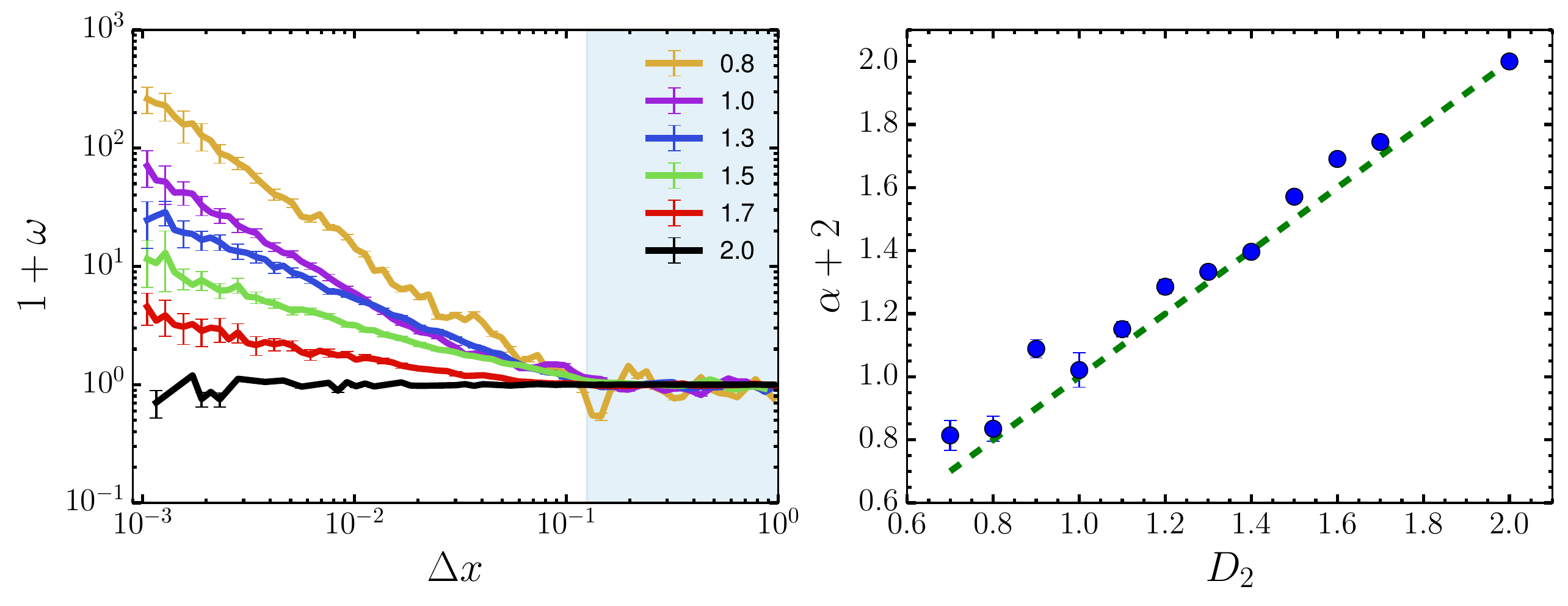}
    \caption{Left: The computed TPCF for fractal models with different input 2D fractal dimensions ($D_2$) as indicated in the legend. The shaded region in blue marks separations beyond the maximum scale of the hierarchy (i.e., $\Delta x >1/2^{l_{\mathrm{base}}}$). As we can see, the TPCF is a pure power law up to scales where the hierarchy extends, beyond which it sharply flattens, and the slope of the power law is progressively shallower at larger $D_2$, with a completely flat TPCF for a purely random distribution (i.e., $D_2=2.0$). Right: Comparison of input fractal dimension provided to the toy model ($D_2$) and the derived 2D fractal dimension $\alpha+2$, obtained from the a least-squares linear fit to the TPCF shown in the left panel at $\Delta x<1/2^{l_\mathrm{base}}$. The one-to-one relation is shown as a dashed green line, and error bars indicate the 1$\sigma$ uncertainties returned by the fit. This shows that the TPCF can reasonably reproduce $D_2$ from the slope $\alpha$ of the power law.}
\end{figure*}

As described, this algorithm is full characterised by the following parameters:
\begin{itemize}
    \item $D_{2}$ : The 2D fractal dimension of the distribution. 
    \item $L_{\mathrm{box}}$: The maximum spatial extent of the box in which the fractal is present.
    \item $l_{\mathrm{base}}$: The minimum level, which consequently sets the maximum separation scale $L_{\mathrm{max}}$ up to which there is fractal structure.
    \item $l_{\mathrm{max}}$: The maximum level, which sets the minimum separation $L_{\mathrm{min}}$ of the hierarchically (fractal) distributed points.
\end{itemize}
Thus, in such a setup, we should expect scale-free behaviour in the ranges of separations from $L_\mathrm{min} \sim L_{\rm box}/2^{l_{\mathrm{max}}}$ to $L_{\mathrm{max}} \sim L_{\rm box}/2^{l_{\mathrm{min}}}$. For the first set of fractal models we prepare, we vary the input fractal dimension $D_{2}$, and use fixed values of $l_{\mathrm{base}} = 2$ and $l_{\mathrm{max}} = 14$, $L_{\mathrm{box}} = 1.0$ for convenience. The TPCFs for the various input fractal dimensions in such a case are shown in Figure~\ref{fig:Fractal_D}. The TPCF is clearly a pure power law up to $L_{\mathrm{max}}$, beyond which it sharply flattens, and the slope of the power law is progressively shallower at larger $D_2$, with a completely flat TPCF for a purely random distribution (i.e., $D_2=2.0$). In addition, we verify that the slope obtained from the power-law part of the TPCF for the fractals matches the theoretical prediction, i.e., $D_2 = 2+\alpha$ \citep{Calzetti_1988,Gomez_1993,Larson_1995}. This is shown in the right panel of Figure~\ref{fig:Fractal_D}, where we compare our input fractal dimension to the model $D_2$ with $2+\alpha$, where we determine $\alpha$ by performing a least-squares fit to the data shown in the left panel at $\Delta x < 1/8$. As we can see, the slopes extracted from the power-law TPCFs match the analytical prediction reasonably well..

\begin{figure}
    \centering
    \includegraphics[width=0.48\textwidth]{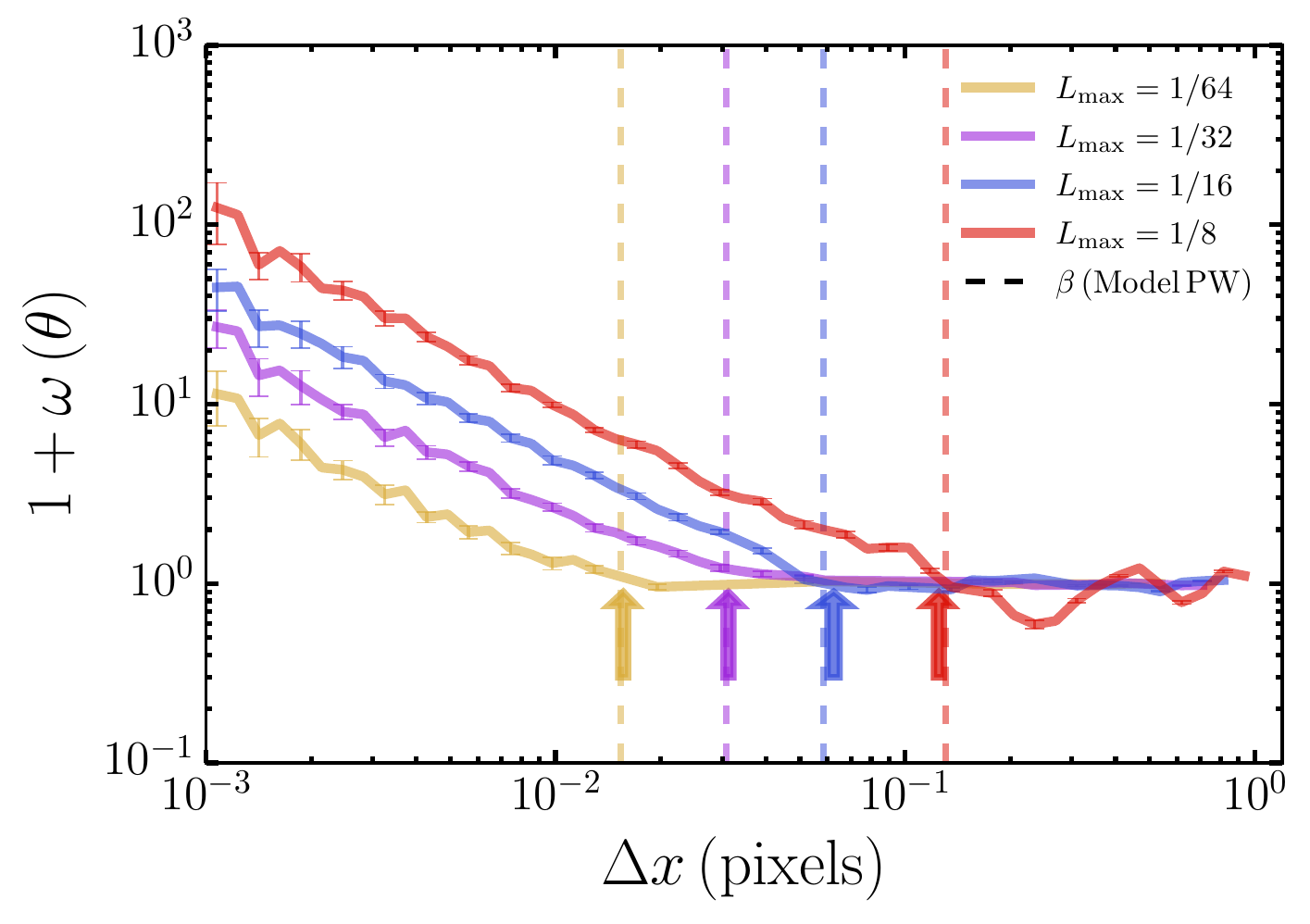}
    \caption{TPCF for fractal models with different input values for the largest scale $L_{\mathrm{max}}$ out to which there is a scale-free hierarchical distribution. The input parameters $D_2 = 1.0$, $L_{\mathrm{min}} = 1/2^{14}$ and $L_{\mathrm{box}} = 1.0$ are kept constant (see Section~\ref{sec:appendix_fractals} for details on the parameters). The solid lines show the TPCFs for the different values of $L_{\mathrm{max}}$, and vertical dashed lines show the inferred value of $\beta$ obtained with a Model~PW fit to the TPCF. Arrows denote the input values of $L_{\mathrm{max}}$ for the four different TPCFs above. It is clear that the best-fit value of $\beta$ in such a scenario traces the largest scale of the scale-free structure $L_{\mathrm{max}}$ quite accurately.}
    \label{fig:Fractal_Lmax}
\end{figure}

In addition to this, we also attempt to vary the maximum scale of the hierarchy $L_{\mathrm{max}}$ in our fractal models, keeping the fractal dimension $D_2$ fixed. This might be important in setting the separation where the TPCF flattens to a value of $1+\omega \sim 1$. This becomes relevant for galaxies whose TPCF is best fit by Model~PW, where the slope of the TPCF changes from a steep one to a relatively much shallower one, characteristic of random distributions. We attempt 4~different values of $L_{\mathrm{max}} = 1/8, 1/16, 1/32$, and $1/64$, keeping $D_2 = 1.0$, $L_{\mathrm{min}} = 1/2^{14}$, and $L_{\mathrm{box}} = 1.0$ fixed. We show the results in Figure~\ref{fig:Fractal_Lmax}. We find that the TPCF flattens more or less at the scale of $L_{\mathrm{max}}$, with a fit to the functional form of Model~PW yielding a transition point $\beta \approx L_{\mathrm{max}}$. This suggests that the transition identified in the power-law behaviour of Model~PW is capturing a physical transition in the underlying distribution from one that is fractal/hierarchical in nature, to a mostly random distribution. 

\subsection{Exponential Disks}
\label{sec:appendix_exponential}
\begin{figure*}
    \centering
    \includegraphics[width=0.98\textwidth]{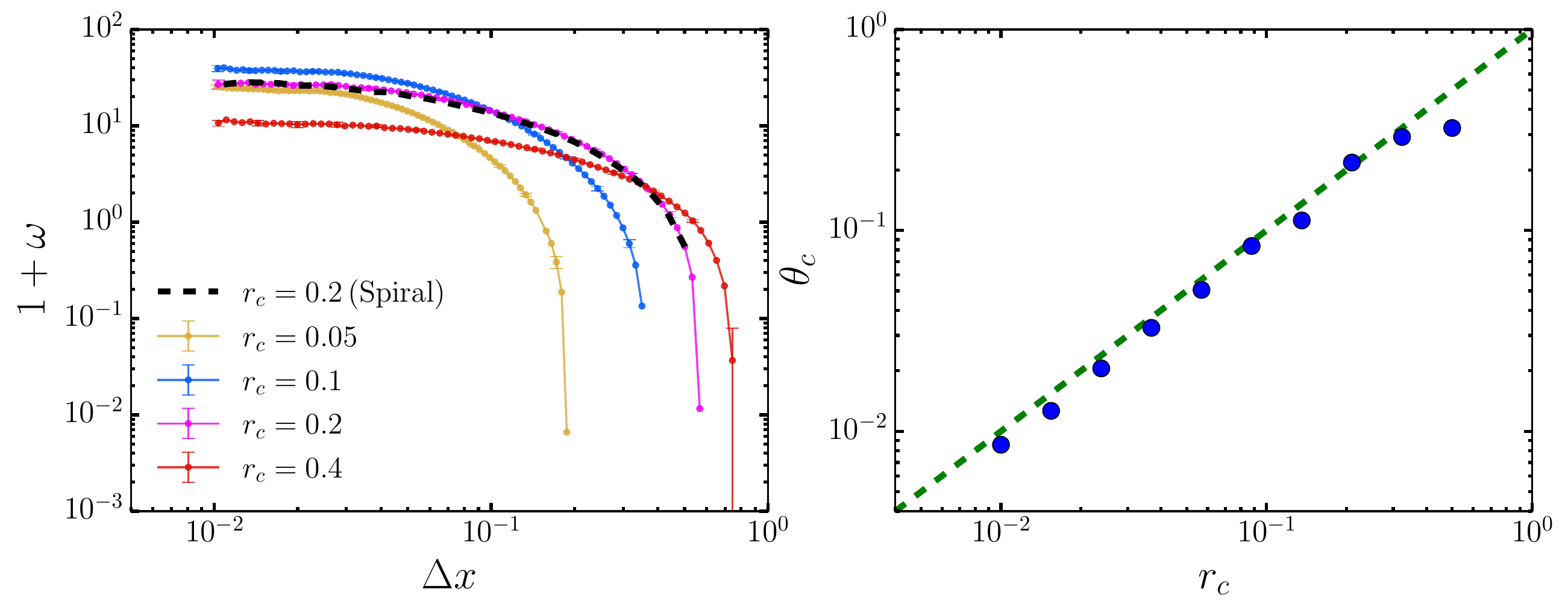}
    \caption{Left: TPCFs obtained with the axi-symmetric thin exponential disk toy models described in Section~\ref{sec:appendix_exponential} for values of $r_c = 0.05,0.1,0.2,$ and $0.4$. The other model parameters are kept fixed at $z_h,i,R_{\mathrm{max}},r_{\mathrm{min}} = 0.2,0.05,30 \degr, 1.0, 0.01$. We also show the TPCF for an exponential disk with $r_c = 0.2$ containing logarithmic spiral arms (black dashed), and find that it is more or less identical to that of an axisymmetric disk, apart from a slight excess of correlation at the smallest separations. Right: The value of $\theta_c$ we obtain by performing a least-squares fit of the functional form for Model~PF (Equation~\ref{eq:ModelPF}) to the measured TPCFs for a range of exponential disc scale radii $r_c$. A one-to-one relation is plotted to guide the eye. We find that the parameter $\theta_c$ reproduces the underlying $r_c$ of the distribution quite well.}
    \label{fig:exponential_TPCF}
\end{figure*}

Here we describe the toy models we use to represent the large-scale distribution of star clusters in thin, radially-exponential disc inclined at an arbitrary angle relative to the line of sight. Our model contains five parameters:
\begin{itemize}
    \item $r_c$: The exponential scale radius of the distribution
    \item $z_h$: The Gaussian scale height of the galaxy
    \item $i$: Line-of-sight inclination angle of the galaxy
    \item $R_{\mathrm{max}}$: The maximum radial extent of the galaxy up to which the points are distributed
    \item $r_{\mathrm{min}}$: The minimum radius at which points can be found from the centre of the galaxy
\end{itemize}
Given these parameters, the model probability density is
\begin{equation}
    \begin{array}{ll}
         & P(r) = \frac{1}{r_c}\exp \left( -\frac{r}{r_c} \right) \; \forall \, r_{\mathrm{min}}\leq r \leq R_{\mathrm{max}}, \\
         & P(z) = \mathcal{N}(0,z_h), \\
         & P(\theta) = \mathcal{U}(0,2\pi),
    \end{array}
\end{equation}
where $r,z,\theta$ are the coordinates of a cylindrical coordinate system with its origin at the galaxy centre and the galaxy midplane lying at $z=0$, $\mathcal{N}(\mu,\sigma)$ is the Gaussian distribution with mean $\mu$ and standard deviation $\sigma$, and $\mathcal{U}(a,b)$ is a uniform distribution in the range $(a,b)$. We generate our galaxy model by drawing $(r,z,\theta)$ coordinates from this distribution, rotating the positions of the points by the chosen inclination angle $i$, and then de-projecting to obtain the plane-of-sky distribution exactly as we do for observed star clusters (see Section~\ref{sec:inclination_correction}). We show the resulting TPCFs for a range of values of $r_c$ in Figure~\ref{fig:exponential_TPCF}; we do not show results for varying $z_h$, because we find that the value of this parameter is negligible as long as $z_h \ll r_c$. The general shape of the TPCFs is a shallow power law at separations $\Delta x \ll r_c$, followed by an exponential fall-off as separations approach $\Delta x \sim r_c$, which is why an exponentially-truncated disc is our prime candidate to describe the Model~PF fits we obtained in Section~\ref{sec:fitting}. We also attempt to fit a Model~PF functional form to the TPCFs of our toy models, and find that it fits very well, with the best-fit value $\theta_c$, i.e., the fitted exponential scale of the fall-off in the TPCF, reproducing the underlying $r_c$ quite well. We demonstrate this in the right-hand panel of Figure~\ref{fig:exponential_TPCF}. In addition, we also attempted introducing an azimuthal pattern, such as a logarithmic spiral, to our exponential disc models. However, we found that the TPCF is relatively insensitive to the introduction of spiral arms, as we show in Figure~\ref{fig:exponential_TPCF}, apart from a slight excess at the smallest separations. 

\begin{figure}
    \centering
    \includegraphics[width=0.48 \textwidth]{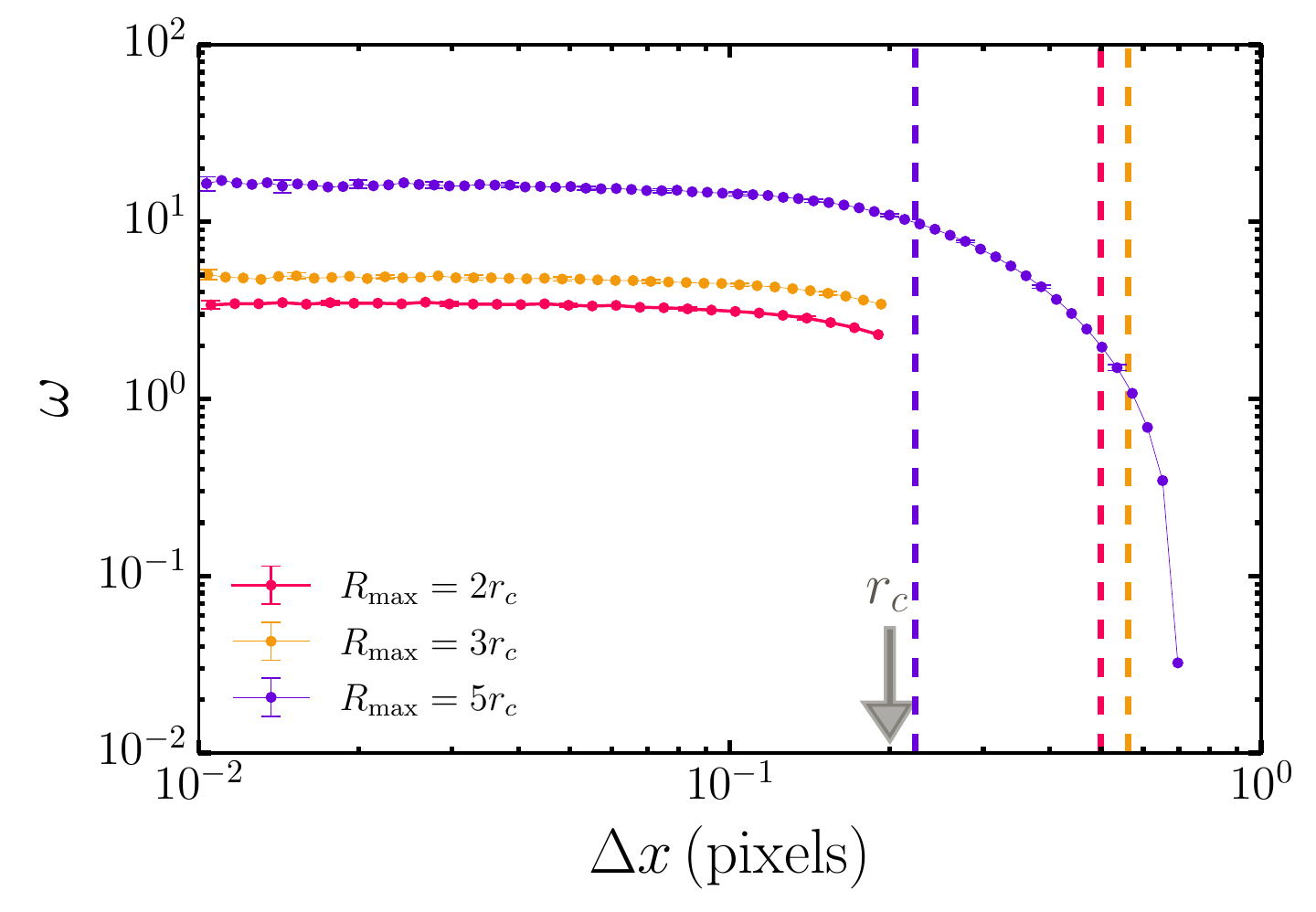}
    \caption{Comparison of the TPCFs (solid lines with error bars) and fits to $\theta_c$ (dashed vertical lines) of galaxy disk models with $r_c,z_h,i,r_{\mathrm{min}} = 0.2,0.05,30 \degr, 0.05$, and 3~different values of $R_{\mathrm{max}} =0.4, 0.6$ and $1.0$, corresponding to $2r_c,3r_c$, and $5r_c$, respectively. As we can see, the fitted value of $\theta_c$ for the latter case is reasonably close to the true value of $0.2$, whereas $\theta_c$ ($\sim 0.6$) for the former two cases overestimates the true value by a factor $\sim 3$. Thus, insufficient radial sampling of the galaxy can lead to an overestimated value for the scale length inferred from the TPCF using Model~PF.}
    \label{fig:Rmax_fits}
\end{figure}

The TPCFs for this model also depend weakly on the other parameters of the toy model. We will not discuss these variations further, except to note that the dependence on $R_{\mathrm{max}}$ becomes relevant to the discussion in Section~\ref{sec:Scale_Radii}, where we find that the inferred scale radii from the Model~PF fit to $\theta_c$ for the larger spiral galaxies is overestimated by a factor \mbox{$2$--$3$} as compared to other estimates in the literature. We understand this to arise due to the fact that a smaller extent of the entire radial distribution of the clusters would be sampled by the HST field-of-view for larger galaxies. To test whether this limited field-of-view can lead to an overestimate of $\theta_c>r_c$ when fitting Model~PF, we set up an exponential disk distribution with $r_c,z_h,i,r_{\mathrm{min}} = 0.2,0.05,30 \degr, 0.05$ and 3~different values of $R_{\mathrm{max}} =0.4, 0.6$ and $1.0$, which corresponds to $2r_c,3r_c$, and $5r_c$ respectively. We then calculate their TPCFs and compare the value of $\theta_c$ with the input $r_c$. This analysis is shown in Figure~\ref{fig:Rmax_fits}. As we can see, when $R_{\mathrm{max}} = 5r_c$, the fitted $\theta_c$ is reasonably close to the input $r_c=0.2$, whereas for $R_{\mathrm{max}} = 2r_c$ and $3r_c$, the fitted $\theta_c$ is considerably higher by a factor $\sim 3$. This shows that if a galaxy with a given scale length $r_c$ is not observed to sufficiently large radii, $r \sim 5r_c$, then the estimate for $\theta_c$ will overestimate the true value of $r_c$ by factors of a few.

\section{Toomre Length Calculation Sources}
Here we list the values and sources for the physical quantities we used, namely the galaxy-averaged gas surface density $\Sigma_\mathrm{g}$ and flat rotational velocity $v_{\mathrm{rot}}$, in the calculation of the average Toomre length in a galaxy $l_{\mathrm{toomre}}$ using Equation~\ref{eq:toomre_length}. For $\Sigma_\mathrm{g}$, we use surface densities of molecular gas as it is the phase of the ISM where star formation is expected to occur \citep{Bigiel_2008}. For $v_{\mathrm{rot}}$, we use \ion{H}{i} rotation curves available in the literature, as this is the most widely available line that traces the rotational velocities in spiral galaxies. We list the values and references for the galaxies below. Note that in some cases where the source does not report a value of $\Sigma_\mathrm{g}$, we explicitly calculate $\Sigma_\mathrm{g}$ by averaging the total molecular gas mass $M_{\mathrm{H2}}$ reported in the source in a disk of radius $R_{25}$, using the values of $R_{25}$ given in Table~\ref{tab:galaxyinfo}. In addition, $l_{\mathrm{toomre}}$ is only computed for the spiral galaxies in our sample.

$\Sigma_\mathrm{g}$: Direct estimate: NGC~0628, NGC~5194, NGC~5457 and NGC~6503 from \citet{Kennicutt_1998}. Indirect calculation: $M_{\mathrm{H2}}$ of NGC~3344 and NGC~3627 from \citet{Young_1989}, $M_{\mathrm{H2}}$ of NGC~1566 from \citet{Bajaja_1995}, and $M_{\mathrm{H2}}$ of NGC~7793 from \citet{Israel_1995}.

$v_{\mathrm{rot}}$: NGC~0628, NGC~3627 and NGC~5194 \citep[THINGS survey, ][]{deBlok_2008}, NGC~1313 and NGC~7793 \citep[Local Volume \ion{H}{i} survey, ][]{Wang_2017,Koribalski_2018}, NGC~1566 \citep[WALLABY, ][]{Elagali_2019}, NGC~3344 \citep{Meidt_2009}, NGC~5457 \citep{Guelin_1970}, NGC~6503 \citep{Greisen_2009}. 


\bsp	
\label{lastpage}
\end{document}